\documentclass{iopart}
\usepackage[square]{harvard}
\usepackage{times}
\usepackage{textcomp}
\usepackage[tight]{units}
\usepackage{graphicx}
\usepackage[utf8]{inputenc}
\usepackage{ifthen}
\ifthenelse{\isundefined{\ead}}
    {
    \usepackage[DIV=15]{typearea}
    }{}

\newcommand{\be}{\begin{eqnarray}}
\newcommand{\ee}{\end{eqnarray}}
\newcommand{\sz}[1]{\sigma^{(z)}_{#1}}
\newcommand{\bra}[1]{\left\langle #1 \right|}               
\newcommand{\ket}[1]{\left| #1 \right\rangle}               

\begin{document}
\ifthenelse{\isundefined{\ead}}
    {
\title{Quantum Simulations with Cold Trapped Ions}
\author{Michael Johanning}
\affiliation{Fachbereich Physik, Universit\"at Siegen, 57068 Siegen,
Germany}
\author{Andr\'{e}s F. Var\'{o}n}
\affiliation{Fachbereich Physik, Universit\"at Siegen, 57068 Siegen,
Germany}
\author{Christof Wunderlich}
\affiliation{Fachbereich Physik, Universit\"at Siegen, 57068 Siegen,
Germany}

    }
    {
\topical{Quantum Simulations with Cold Trapped Ions}
\author{Michael Johanning, Andr\'{e}s F. Var\'{o}n, Christof Wunderlich}
\address{Fachbereich Physik, Universit\"at Siegen, 57068 Siegen,
Germany}

\ead{wunderlich@physik.uni-siegen.de}
\submitto{\jpb}
    }

\begin{abstract}
The control of internal and motional quantum degrees of freedom of
laser cooled trapped ions has been subject to intense theoretical
and experimental research for about three decades. In the realm of
quantum information science the ability to deterministically prepare
and measure quantum states of trapped ions is unprecedented. This
expertise may be employed to investigate physical models conceived
to describe systems that are not directly accessible for
experimental investigations. Here, we give an overview of current
theoretical proposals and experiments for such quantum simulations
with trapped ions. This includes various spin models (e.g., the
quantum transverse Ising model, or a neural network), the
Bose-Hubbard Hamiltonian, the Frenkel-Kontorova model, and quantum
fields and relativistic effects.

\end{abstract}


 \maketitle


\section{Introduction}

Atomic ions confined in an electrodynamic trap provide us with
individual quantum systems whose internal and external degrees of
freedom can be controlled to a considerable degree by the
experimenter. A strong motivation for early experiments with
individual trapped ions arose from their potential use as frequency
standards (e.g., \cite{Dehmelt1981} and references therein). Today,
by using trapped ions as frequency standards unsurpassed accuracy
and precision has been reached (e.g., \cite{Rosenband2008} and
references therein), and in addition, trapped ions are used for a
wide range of investigations into fundamental questions of physics,
for instance in the field of quantum information science. This
special issue gives an account of the many facets of research with
trapped ions.

Controlling motional and internal states of trapped ions by letting
them interact with electromagnetic fields, ranging from uv light to
rf radiation, has lead to the realization of experiments that
previously (reaching well into the second half of the 20th century)
were only conceivable, by most physicists, as Gedankenexperiments. A
small (and unsystematic) sample from such fascinating experiments
include the trapping and visualization of a single ion
\cite{Neuhauser1980}, the laser cooling of a single ion to its
motional ground state \cite{Diedrich1989}, the deterministic
generation and analysis of Schr\"odinger cat and other entangled
states of massive particles,
\cite{Monroe1996,Leibfried2005,Haeffner2005,Blatt2008}, or the
teleportation of ions \cite{Riebe2004,Barrett2004}.

While building a universal quantum computer still poses formidable
experimental challenges, in particular because of the low error
thresholds required for fault-tolerant computing it appears that
useful quantum simulations are more amenable to experimental
efforts.

There are two common usages of the term "quantum simulation": (i)
the simulation of static and dynamic properties of quantum systems
(e.g. atoms, molecules, condensed matter systems) using {\em
classical} computers, and (ii) the simulation of the properties of
one quantum system by means of controlling and observing another
quantum system. Here, we are concerned with the second type of
research.

The purpose of a quantum simulation is to use a well-understood
physical A system to simulate the static properties and the dynamics
of another system B that is difficult or, for all practical
purposes, even impossible to investigate experimentally
\cite{Feynman1982,Feynman1986,Jane2003}. In order to be able to
simulate B by observing A, both physical systems need to be
described by the same underlying mathematical model.

In this article we give an overview of concrete proposals for
performing experimental quantum simulations with laser cooled
trapped ions, and of experimental work immediately relevant for
using trapped ions as a quantum simulator.

\section{Spin models}
Interacting spins play an important role in many physical models in
a variety of research fields of physics, for instance, in models
that are employed to describe phenomena in condensed matter physics
(e.g., \cite{Schollwoeck2004} and references therein). Also, in
recent years the connection between entanglement on the one hand,
and static and dynamic properties of many-body systems on the other
hand, for instance, near quantum phase transitions has attracted
much interest (e.g., \cite{Osterloh2002,Amico2008,Plenio2007} and
references therein).

Probably the simplest approach to modeling interacting spins is the
classical Ising model \cite{Lenz1920,Ising1925} where the energy of
a system of localized particles is given by
 \be
 E = - J \sum_{\langle i>j \rangle} s_i s_j
 \label{eq:Ising}
 \ee
and each particle (spin) maybe in two possible states indicated by
$s_i = \pm 1$. The summation includes the interaction between
nearest neighbours only, indicated by $\langle i>j \rangle$. Beyond
interacting spins, the Ising model is often used to approximate
systems where the constituents may be in one of two possible states,
describing, for example, the tunneling between two potential wells
or the occupation of a lattice site with two types of particles
\cite{Brush1967}. Ising already showed that a phase transition
between an ordered state  and a disordered state does not take place
in this model in one dimension at non-zero temperature, whereas in
two dimensions such a phase transition is possible as shown by
Onsager \cite{Onsager1944}. However, extending the range of
interaction beyond nearest neighbours gives rise to phase
transitions even in one dimension \cite{Gitterman2004}.

A more sophisticated model of the interaction between localized
spins \cite{Heisenberg1928} that, in addition, are exposed to an
external local magnetic field $\vec{B}_i$ uses the Hamiltonian
 \be
 H_{s} = - \frac{2}{\hbar}\sum_{i>j} J_{ij} \vec{S}_i \cdot \vec{S}_j - \frac{g
 \mu_B}{\hbar} \sum_i \vec{S}_i \cdot \vec{B}_i \ .
 \label{eq:H_spin}
 \ee
Here we assume that all lattice sites are populated by particles
with spin $\vec{S}_i$ with identical $g$-factor and eigenvalues of
$S_z$ that are multiples of $\hbar/2$ ($\mu_B$ is the Bohr
magneton). No assumption has yet been made on the physical nature
and range of the spin-spin coupling which often is caused by a
dipole-dipole or exchange interaction. Therefore, it is useful to
explicitly parametrize a possible anisotropy of this coupling by
introducing the constants $J_{ij}^{(\alpha)}$, with $\alpha=x,y,z$
so that
 \be
 H_{s}^{(a)} = - \frac{2}{\hbar}\sum_{\alpha=x,y,z}\sum_{i>j} J_{ij}^{(\alpha)} S_i^{(\alpha)} S_j^{(\alpha)}
 -  \sum_i \vec{S}_i \cdot \vec{B}_i' \ ,
 \label{eq:H_spin_an}
 \ee
where the scaled local fields are defined as $\vec{B}_i' \equiv
\frac{g \mu_B}{\hbar} \vec{B}_i$.

When describing physical phenomena that occur in condensed matter,
for example, concerning magnetism, this Hamiltonian is often subject
to idealizations with regard to its dimensionality, the range of
interaction, its (an-)isotropy, the uniformity of coupling constants
$J_{ij}$, the magnitude and direction of additional (local or
global) fields, and regarding the finite size of a particular system
under consideration.

The Hamiltonian of the quantum transverse Ising model is obtained
from $H_{s}^{(a)}$ for spin-1/2 particles (i.e., $\vec{S} =
(\hbar/2) \vec{\sigma}$) by setting $ J_{ij}^{(x)}= 0 =
 J_{ij}^{(y)}$ and then including further restrictions:
Specifically, the local fields $\vec{B}_i$ are replaced by a global
field $\vec{B}=(B,0,0)^T$ pointing in the $x-$direction, thus
neglecting inhomogeneities in the physical system to be modeled
\cite{Igloi1993,Platini2007}. Additional approximations often made
are that the interaction between spins is restricted to nearest
neighbours, and uniform coupling $J_{ij}^{(z)} = J \forall i, j$ is
assumed. This then leads to
 \be
  H_{TI} =  - \frac{\hbar}{2} J \sum_{\langle i>j\rangle} \sigma_i^z \sigma_j^z
  - B_x' \sum_i \sigma_i^x \ .
 \label{eq:TI}
 \ee
The quantum transverse Ising model is a prominent example for a spin
model \cite{Pfeuty1970} that has been the starting point for many
theoretical investigations (e.g., \cite{Das2005} Part I and
references therein).

\subsection{Simulation of spin models with trapped ions}

Using two internal states of a trapped ion as an effective spin-1/2
with the ions interacting via the Coulomb force allows for
engineering spin couplings by letting the ions interact, in
addition, with external dynamic and static fields. Initial
theoretical and experimental work was mainly concerned with creating
effective interactions between two spins to be used as 2-qubit
quantum gates
\cite{Cirac1995,Monroe1995,Molmer1999,Milburn1999,Milburn2000,Sorensen2000,Jonathan2000,Jonathan2001,Cirac2000,Sasura2003,Schmidt-Kaler2003,Leibfried2003B}

\subsubsection{Direct spin-spin interaction}
In this section we concentrate on couplings of the type given in
equation \ref{eq:H_spin_an} that, for example, allow to study a
phase transition between ordered and disordered phases of an ion
crystal as will be outlined in section \ref{sec:Phase}. The question
we want to consider first is, How can such a spin-spin coupling be
obtained in an ion trap and how can it be engineered to simulate a
specific Hamiltonian?

If we consider a laser cooled Coulomb crystal of singly charged ions
confined in a linear trap such that they arrange themselves in a 1-d
string, then the relevant length scale for the inter-ion distance is
 \be
 \zeta\equiv (e^2/4\pi\epsilon_0 m \nu_1^2)^{1/3} \ ,
 \ee
where $m$ is the mass of one singly charged ion, $e$ the elementary
charge, and $\nu_1$ is the angular vibrational frequency of the
center-of-mass axial mode of the collection of ions
\cite{Steane1997,James1998}. We have $\zeta=1.5 \times 10^{-5}
(Mf^2)^{-1/3}$~m with $M$ in atomic mass units and $f = \nu_1/2\pi$
in units of MHz. Thus, $\zeta$ typically amounts to a few
micrometers. For a linear ion string the inter-ion spacing is
$\delta\!z \approx \zeta\, 2N^{-0.56}$ \cite{Schiffer1993,Dubin1993}
which gives 3.7~$\mu$m for $^{171}$Yb$^+$ and 9.9~$\mu$m for
$^{9}$Be$^+$ assuming two ions confined in a potential characterized
by $f=1$ MHz.

In principle, a direct spin-spin interaction may take place via a
magnetic moment associated with the (pseudo-)spin of the ions or via
an exchange interaction. However, as will be recapitulated now, the
inter-ion spacing renders this interaction negligible. It is
nevertheless instructive to have a look at the dipole-dipole
interaction, since it may help to gain an intuitive understanding
also of the {\em indirect} spin-spin coupling treated in section
\ref{sec:Spin-spin}.

\begin{figure}[tbh]
    \centering
       \includegraphics[width=\columnwidth]{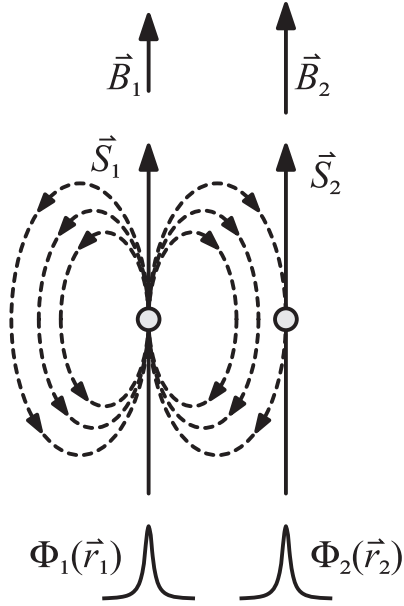}
    \caption{An artist's illustration of the coupling between two spins
via a dipole-dipole interaction. Two classical magnetic moments
$\gamma_{i}\vec{S}_{i}$ fixed in space are exposed to individual
external magnetic fields $\vec{B}_{i}$ ($i=1,2$) that determine
their respective energy for a given orientation (parallel or
anti-parallel to $\vec{B}_{i}$). Spin 2 experiences in the presence
of spin 1, in addition to $\vec{B}_{2}$, an additional field that
depends on the relative orientation between the two spins. Thus the
energy of spin 2 is shifted depending on the orientation of spin 1
(and vice versa). The spatial wavefunctions of the two spins are
indicated by $\Phi_i(\vec{r}_i)$. }
    \label{fig:spinspin}
\end{figure}

\begin{figure}[tbh]
    \centering
        \includegraphics[width=\columnwidth]{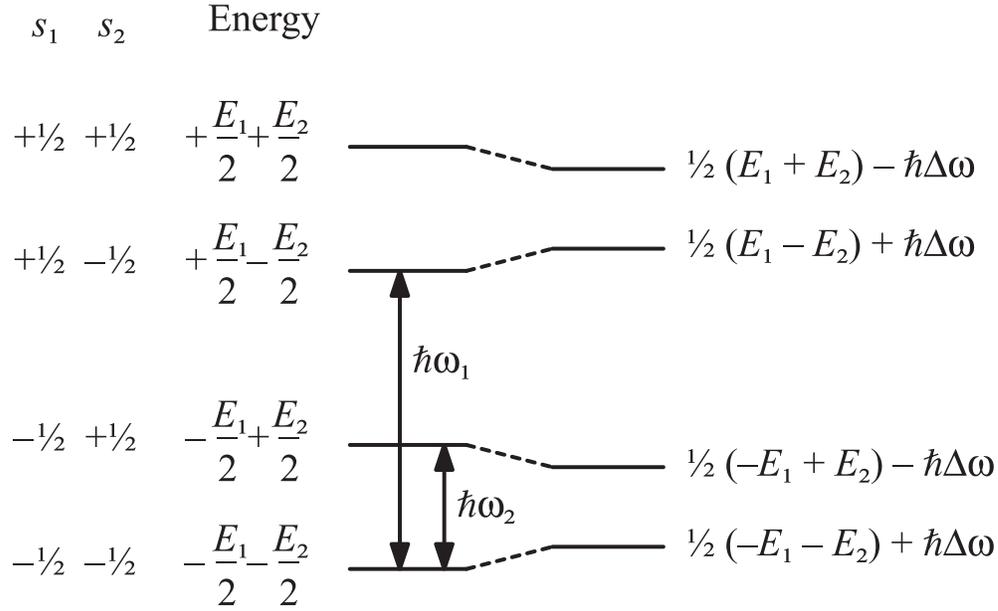}
    \caption{Energy levels of two spins as a function of their orientation.
    The quantum numbers $s_i$, $i=1,2$ indicate the spin state. The resonance
    frequencies of non-interacting spins are $\omega_i$ while when they interact
    these resonances are shifted by $\Delta\!\omega_i$. }
    \label{fig:Elevsls}
\end{figure}

For the moment, for the sake of obtaining an intuitive picture, we
consider two spins with their relative alignment fixed, and exposed
to local magnetic fields $B^{(z)}_{1,2}$ such that the uncoupled
Hamiltonian reads
 $H_0 = (\hbar/2) (\omega_1 \sz{1} + \omega_2 \sz{2})$
($\omega_{1,2}$ are the angular Larmor precession frequencies).
Treating the magnetic dipole-dipole interaction as a perturbation to
first order, its magnitude is determined by the expectation value
 \be
 \bra{s_1, s_2} V_{MD} \ket{s_1, s_2} = s_1 s_2  f_{\theta}(\delta\!z)\equiv
 \pm \hbar \Delta\omega
  \label{eq:V_MD}
 \ee
where $s_{1,2}= \pm 1$ and
 \be
  V_{MD} &=& -\frac{\hbar^2}{4} \frac{\mu_0}{4\pi} \frac{\gamma_1 \gamma_2}{\delta\!z^3}
(3\cos^2(\theta)-1) \sz{1} \sz{2} \nonumber \\
 &\equiv& f_{\theta}(\delta\!z)\sz{1} \sz{2}
 \ .
 \ee
Here, the gyromagnetic ratio of spins 1 and 2 is denoted by
$\gamma_{1,2}$, $\mu_0$ indicates the permeability of empty space,
and $\theta$ is the angle indicating the relative orientation of the
two spins. Equation \ref{eq:V_MD} tells us that the energy of the
two-spin systems depends on the relative orientation between the two
spins, that is, flipping either one of the two spins results in a
change of energy. The physical reason is that the energy of a given
spin depends not only on the imposed local field $B^{(z)}_{1,2}$,
but also on the field generated by the second spin at the location
of the first one, which in turn depends on its orientation (compare
figure \ref{fig:spinspin} and figure \ref{fig:Elevsls}).

For two electron spins ($\gamma_{1,2} \approx e/m_e$ with mass
$m_e$) and assuming $\theta=0$, we find $\Delta\omega \approx
2\pi\times 2.6\times 10^{-20} \delta\!z^{-3}$~Hz with $\delta\!z$
measured in meters. Therefore, for a typical inter-ion separation,
$\delta\! z=5\mu$m the direct magnetic spin-spin interaction amounts
to $\Delta\omega \approx 2\pi\times 0.2$ mHz and is negligible in
typical ion trap settings. This estimate is based on a magnetic
dipole-dipole interaction of ions in their ground state. If bringing
the ions to a highly excited Rydberg state, one would expect an
appreciable interaction between their electric dipoles
\cite{Muller2008,Urban2009,Gaetan2009}.

In order for the exchange interaction to play a non-negligible role,
an overlap of the wavefunctions describing the external, motional
degrees freedom of neighbouring ions would be required, or more
precisely, the exchange integral \mbox{$\int
r_{12}^{-1}[\Phi_n(\vec{r}_1)\Phi_{n+1}(\vec{r}_2)][\Phi_n^*(\vec{r}_2)\Phi_{n+1}^*(\vec{r}_1)]
$} would have to have a non-negligible magnitude. Here, the spatial
wavefunction of ion number $n$ ($n+1$) located at position
$\vec{r}_{1}$ ($\vec{r}_{2}$) is $\Phi_n(\vec{r}_1)$
($\Phi_{n+1}(\vec{r}_2)$) and the integration is carried out over
all space. However, the root mean square extension of the ground
state wavefunction of a trapped ion is given by $\Delta z_1 =
\sqrt{\hbar/2m\nu_1} = 7.1\times 10^{-8}\times (Mf)^{-1/2}$~m In the
case of a $^9$Be$^+$ ($^{171}$Yb$^+$) ion, one obtains $\Delta\!z_1
= 24$ nm (5.4 nm) for $f=1$MHz which typically is three orders of
magnitude smaller than the inter-ion separation $\delta\!z$.
Therefore, the exchange interaction can be neglected, too.

\subsubsection{Spin-spin interaction mediated by the Coulomb force}
\label{sec:Spin-spin}

Thus, a direct spin-spin interaction between trapped ions will not
be sufficiently strong for a useful implementation of the spin
models described above. Instead, we now look for an interaction that
is mediated by some other physical mechanism.

\paragraph{Physical picture}

The applied trapping potential and the Coulomb forces between the
ions determine their equilibrium positions. Now, by applying a
suitable external field, an additional force is exerted on each ion
whose magnitude and/or direction depends on the orientation of the
spin of the ion and varies spatially (i.e., depends on the exact
location of the ion). If this state-dependent force acts in addition
to the trap and Coulomb forces, then a spin flip of ion 1 changes
this force and, as a consequence, this ion changes its equilibrium
position. Via Coulomb repulsion, this in turn leads to a shift of
the equilibrium position of the neighbouring ion 2 that now finds
itself exposed to a field of different magnitude and thus changes
its internal energy.

This additional field may be a static magnetic field with spatially
varying magnitude. The direct interaction between magnetic dipoles
(where flipping one spin amounts to a change in the local magnetic
field experienced by another nearby spin) is replaced by an indirect
interaction where the change in the magnetic field at the location
of spin 2 is brought about by a change of its equilibrium position
(caused by the spin flip of ion 1) in a spatially varying external
field.

For a more quantitative description we consider the force $\vec{F}=
\langle (\vec{\mu} \cdot \vec{\nabla})\vec{B}\rangle$ acting on an
atomic magnetic moment $\vec{\mu}= (\gamma \hbar/2) \vec{\sigma} $
associated with spin $(\hbar/2) \vec{\sigma}$ in a magnetic field
$\vec{B}$. Without loss of generality we restrict the following
discussion to the case where $\vec{B}=\vec{B_0} + b\hat{z}$ with
constant offset field $\vec{B_0}$ and gradient $b$ along the
$z-$axis \footnote{Such a field may be {\em approximately} realized,
even though it does not fulfill $\vec{\nabla}\cdot \vec{B} = 0$, for
example, on the axis of rotational symmetry of a quadrupole field
sufficiently far from the point of inflection symmetry.} such that
the force $F_z = \langle(\gamma \hbar/2) b \sz{} \rangle$ changes
sign upon flipping the atomic spin ($\langle\rangle$ indicates the
expectation value). Using the position dependent spin resonance
frequency $\omega = \gamma b z$, the force can be expressed as a
function of the change of $\omega$ when moving the ion along the
$z-$axis:
 \be
 F_z= (\hbar/2) \partial_z \omega \langle \sz{} \rangle \ .
 \ee

\begin{figure}[tbh]
    \centering
       \includegraphics[width=\columnwidth]{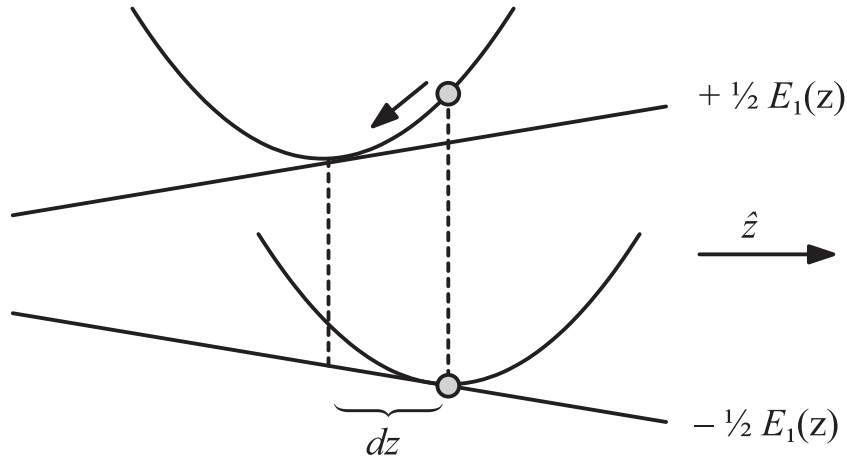}
    \caption{A single ion confined in a potential that is the sum of a harmonic
    oscillator and a linear term. The sign of the linear term depends on the
    internal spin state ($\pm \frac{1}{2} E_1$) of the ion. Upon flipping the ion's spin
     its motion can be excited.}
    \label{fig:coupling}
\end{figure}

For an ion of mass $m$ confined in a harmonic potential,
characterized by angular frequency $\nu$, this linear force that
acts in addition to the trapping potential and Coulomb force, shifts
the equilibrium position of the ion, upon flipping its spin, by an
amount $d_z= \mp F_z /(m\nu^2)$ from its initial equilibrium
position (compare figure \ref{fig:coupling}). Thus, in a classical
picture, the ion finds itself on the slope of a harmonic oscillator
potential, instead of at the potential minimum, after its spin was
flipped. Consequently, it will start to oscillate around the new
equilibrium position. In this way the ion's internal dynamics and
its motional degrees of freedom can be coupled, even if the
radiation causing the spin flip doesn't impart enough linear
momentum to excite the ion's motion. This coupling, quantum
mechanically described by an effective Lamb-Dicke parameter
$\eta_{eff}= d_z / \Delta\!z$
\cite{Mintert2001,Mintert2001E,Wunderlich2002,Wunderlich2003}, has
recently been observed experimentally \cite{Johanning2009}.

\begin{figure}[tbh]
    \centering
        \includegraphics[width=\columnwidth]{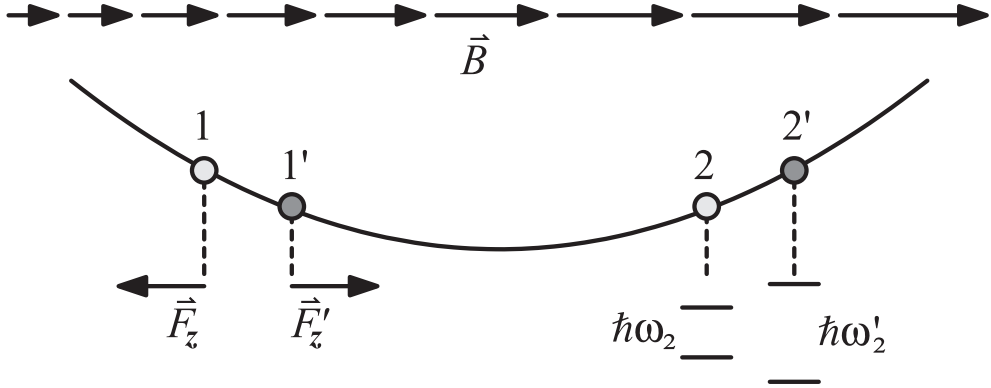}
    \caption{Sketch of two ions confined in a harmonic potential with superimposed
    spatially varying magnetic field $\vec{B}$ to illustrate spin-spin coupling
    mediated by Coulomb repulsion. When flipping the spin of ion 1,
    the state-dependent force acting on it will change sign, and thus, its equilibrium
    position will change ($1 \rightarrow 1'$). Through Coulomb interaction the second ion, too,
    will change its equilibrium position ($2 \rightarrow 2'$), and, since it moves in
    a field gradient, consequently will change its internal resonance frequency from $\omega_2$ to
    $\omega_2'$.}
    \label{fig:gradient}
\end{figure}

In order to see how the desired $J-$coupling arises, we consider how
an ion's shift of its equilibrium position affects its internal
energy. This change of its internal energy is given by
 \be
  \hbar J= - F_z d_z= - F_z^2 /(m\nu^2) \propto (b/\nu)^2.
 \label{eq:J_simple}
 \ee
As a consequence of one ion's shift of equilibrium position, through
Coulomb interaction, the position and energy of neighbouring ions is
shifted, too, giving us the desired (indirect) spin-spin interaction
(compare figure \ref{fig:gradient}).

\paragraph{Calculation of spin-spin coupling}

After having introduced this general physical mechanism that may be
used to induce an effective spin-spin interaction, we will now
briefly outline how this may be employed to engineer a variety of
interactions between a collection of trapped ions. We consider the
internal energies of $N$ ions all of which experience a shift of
their internal resonance frequency that, for now, depends on their
axial position denoted by $z$, only:
 \be
 H_I = - \frac{\hbar}{2} \sum_{n=1}^N \omega_n \sigma_z^{(n)} -
\frac{\hbar}{2} \sum_{n=1}^N \partial_z\omega_n \, q_n
\sigma_z^{(n)}
 \label{eq:H_i}
 \ee
Here, $q_n$ is the deviation of ion $n$ from its equilibrium
position $z_n^{(0)}$,
 \be
  \omega_n \equiv \omega\Big|_{z_n^{(0)}} \qquad ,  \qquad
  \partial_z\omega_n \equiv \partial_z \omega \Big|_{z_n^{(0)}}
  \ ,
 \ee
and the expansion of $\omega_n$ has been included up to first order
\footnote{We assume a constant field gradient, that is, second and
higher order derivatives of the field vanish. However, the
expression \ref{eq:H_i} is still valid even if these derivatives are
different from zero as long as the {\em gradient} doesn't vary
appreciably over the extent $\Delta\!z$ of an ions spatial
wavefunction.}.

The external (motional) degrees of freedom of $N$ ions are
determined by the applied trapping potential and their mutual
Coulomb repulsion. Expanding the total potential up to second order
gives
 \be
V^{(harm)} = \frac{m}{2} \sum_{\alpha=1}^3 \sum_{i,j=1}^N
A_{\alpha\alpha,ij} \, q_{\alpha,i} \, q_{\alpha,j}
 \ee
where the three spatial degrees of freedom $\alpha = x,y,z$ are
uncoupled \cite{James1998}. The Hessian matrices
$A_{\alpha\alpha,ij}$ that characterize the potential experienced by
the ions are real valued and symmetric in $i$ and $j$ and are
transformed into diagonal matrices $D_{\alpha}$ via $\label{def_S}
D_{\alpha} = S_{\alpha}^{-1} \, A_{\alpha}^{(2)} \, S_{\alpha}$ with
eigenvectors $\vec{Q}$.

In terms of the normal coordinates we obtain the following
Hamiltonian describing $3N$ uncoupled harmonic oscillators
\cite{James1998}:
 \be
 H_{E} = \sum_{\alpha=1}^3 \left[
\frac{1}{2m} \sum_{n=1}^N P_{\alpha,n}^2 +
   \frac{m}{2} \sum_{n=1}^N \nu_{\alpha,n}^2 \, Q_{\alpha,n}^2 \right]
 \label{eq:H_e}
 \ee
where $\nu_{\alpha,n}$ is the angular frequency of the normal
vibrational mode $n$ in direction $\alpha = x,y,z$ and the relation
between the normal coordinates $Q_{\alpha,n}$ and the local
coordinates $q_{\alpha,n}$ is $\vec{Q}_{\alpha} \equiv
(Q_{\alpha,1},\ldots,Q_{\alpha,N}) = S_{\alpha}^{-1} \,
\vec{q}_{\alpha}$ with $\vec{q}_{\alpha} =
(q_{\alpha,1},\ldots,q_{\alpha,N})$, furthermore $P_{\alpha,n}
\equiv m \, \dot{Q}_{\alpha,n}$.

The total Hamiltonian includes internal and motional parts $H = H_I
+ H_E^{(z)}$, where, for now, we restrict our considerations to the
spatial $z-$direction where a field gradient is present. If a
magnetic field gradient was present also in the radial direction
(which is the case, e.g., for a quadrupole field close to its point
of symmetry), then the radial vibrational modes would also
contribute to the $J-$coupling. However, this contribution usually
is small, since in order to maintain a linear ion string, the
angular frequencies $\nu_{x,1}$ and $\nu_{y,1}$, characterizing the
radial trapping potential, are usually chosen much larger than
$\nu_{z,1}$. Taking into account that the $J-$coupling is
proportional to the inverse square of the trap frequency (taking
equation \ref{eq:J_simple} as a guideline), we neglect the
contribution from transverse vibrational modes for now. In
\cite{Porras2004B,Deng2005} transverse modes are explicitly
considered (see below).

After expressing the local coordinates $q_n $ in equation
\ref{eq:H_i} by the normal coordinates, $q_n=\sum_{l=1}^N
S_{nl}Q_l$, noting that $Q_l= \Delta\!z_l(a^\dagger_l + a_l), P_l=i
m \nu_l \Delta\!z_l(a^\dagger_l - a_l)$, and performing a suitable
unitary transformation of $H$, $\tilde{H}= U^\dagger H U$, with
 \be
 U \equiv \exp\left( -\frac{i}{\hbar}\sum_{n,l=1}^N \Delta\!z_l\varepsilon_{nl} \sigma_z^{(n)} P_l \right).
 \ee
we obtain \cite{Wunderlich2002}
 \be
 \nonumber
\tilde{H} = - \frac{\hbar}{2}\sum_{n=1}^N \omega_n\sigma_z^{(n)} +
\hbar\sum_{n=1}^N \nu_n a_n^{\dagger} a_n
      - \frac{\hbar}{2}\sum_{i,j=1 \atop i<j}^N J_{ijl} \sigma_z^{(i)} \sigma_z^{(j)} \; .
 \label{eq:H_tilde}
 \ee
where the vibrational motion and the internal states are now
decoupled ($a_n^{\dagger}, a_n$ are the creation and annihilation
operator of vibrational mode $n$). The spin-spin coupling constants
are given by
 \be
 J_{ij} \equiv \sum_{n=1}^N \nu_n \varepsilon_{in}
 \varepsilon_{jn}
 \label{eq:J}
 \ee
where the dimensionless constants
 \be
 \varepsilon_{in} \equiv
 \frac{\Delta z_n \partial_z\omega_i}{\nu_n} S_{in}
 \label{eq:eps}
 \ee
have been introduced. The numerator of the fraction on the
right-hand-side of equation \ref{eq:eps} gives the change of the
internal energy (or frequency) that ion number $i$ undergoes when
its position is shifted by an amount equal to the extension of the
ground state wavefunction, $\Delta z_n=
\sqrt{\frac{\hbar}{2m\nu_n}}$ of vibrational mode $n$. This is set
in relation to the energy (frequency) of a phonon  of mode $n$. The
matrix element $S_{in}$ gives the scaled deviation of ion $i$ from
its equilibrium position when vibrational mode $n$ is excited. Thus,
$\varepsilon_{in}$ tells us how strongly ion number $i$ couples to
vibrational mode $n$ when its spin is being flipped. The sum in
equation \ref{eq:J} extends over all vibrational modes, since the
local deviation, $q_n$ of ion $i$ from its equilibrium position
potentially (if $S_{in}$ is different from zero) excites all
vibrational modes of the ion string.

Noting that $\nu_n \propto \nu_1$ and $F_{z,i} \propto
\partial_z\omega_i$ and plugging the expression for $\Delta z$ into
equation \ref{eq:eps} we find from equation \ref{eq:J} that
 \be
 J_{ij} &\propto& \frac{F_{z,i}F_{z,j}}{\nu_1^2} \nonumber \\
        &\propto& \left( \frac{b}{\nu_1} \right)^2
 \label{eq:J_prop}
 \ee
in agreement with the result \ref{eq:J_simple} that was found from a
simple model based on classical considerations (here, $\nu_1$ is the
angular frequency of the centre-of-mass mode of $N$ ions). Going
from the first to the second line in equation \ref{eq:J_prop}, we
have assumed a force caused by a constant magnetic field gradient
causing a linear Zeeman shift of the spin states of ions $i$ and
$j$.

This spin-spin coupling is only weakly sensitive to thermal
excitation of the ion string. Thus, usually simple Doppler cooling
will be sufficient to avoid unwanted thermal effects on the coupling
constants $J_{ij}$ \cite{Loewen2004}. Detailed calculations on this
aspect will be published elsewhere.

In \cite{Ospelkaus2008} it is shown how an {\em oscillating}
magnetic field generated in a micro-structured surface trap may be
used to induce spin-coupling between ions and how this could be
employed to carry out quantum gates.

\paragraph{Spin-spin interaction created by an optical force}

The $J-$coupling described above may also be created by a
state-dependent {\em optical} force (instead of the magnetic force
assumed above) induced by an off-resonant standing laser wave
\cite{Porras2004B,Deng2005}. Applying state-dependent forces (e.g.,
optical or magnetic forces) along different spatial directions,
combined with various trapping conditions, a variety of different
coupling Hamiltonians of the general form
 \be
 H_{s}^{(ion)} = - \frac{1}{2}\sum_{\alpha=x,y,z}\sum_{i>j} J_{ij}^{(\alpha)}
 \sigma_i^{\alpha} \sigma_j^{\alpha}
 -  \sum_i \sigma_i^{\alpha} \vec{B}_i^{\alpha '} \
 \label{eq:H_s_ion}
 \ee
may be realized \cite{Porras2004B,Deng2005}.

In particular, in \cite{Deng2005} it is shown how state-dependent
forces that act in the radial direction of a 1-d ion string may be
exploited to simulate a transverse quantum Ising model and an
XY-model, either with Coulomb crystals confined by a global
potential, or linear arrays of microtraps where each ion is confined
in its own, externally controlled  potential well. It is
demonstrated (by means of numerical and analytical calculations)
under which conditions quantum phase transitions should be
observable with 1-d strings of trapped ions, and what the expected
characteristic features of these transitions are, as compared to the
usual Ising and XY-models where nearest-neighbour interaction only
is considered.

\subsubsection{Engineering the spin-spin coupling}

A straight forward way to apply a magnetic gradient that induces the
long-range $J-$coupling described above in a linear Coulomb crystal
is the use of current carrying coils in an
anti-Helmholtz type arrangement, mounted such that their axis of
rotational symmetry coincides with the trap $z-$axis. The field
generated by these coils is easily calculated analytically and
gradients up to several tens T/m are attainable with coils with a
diameter of typically 1 mm. Some examples for possible
configurations and computed examples for $J-$coupling constants are
presented in \cite{Wunderlich2003,Wunderlich2005,Wunderlich_H2009}.
Such coils may also be integrated into micro-structured ion traps
that are currently being developed. Here, larger gradients are
realistic (several hundreds of T/m) \cite{Brueser2009}. When using
current-carrying structures for generating magnetic fields, the
currents, and thus the $J-$coupling, can be turned on and off as
required, however, the stability of the magnetic field needs
attention to avoid dephasing of spins. Alternatively, permanent
magnets may be used \cite{Johanning2009}.

\paragraph{Global trap potential}

If the ions are confined in one common trapping potential, then the
strength of the $J-$coupling may be controlled globally, that is,
all coupling constants may be varied together by adjusting the
gradient $b$ and/or the secular trap frequency $\nu_1$. In a linear
Paul trap the trap frequency along the $z-$axis of a linear trap is
adjustable by applying a suitable voltage to the end cap electrodes,
while the radial trapping frequency can be adjusted by varying the
amplitude of the applied rf trapping field. If the gradient $b$ is
generated by electrical currents, then these currents can be varied,
and turning the magnetic field on and off using a desired envelope
allows for simultaneous switching of all coupling constants. If
permanent magnets are used, and their position is fixed along the
$z-$axis (typically generating a quadrupole field), the ions may be
moved collectively along the $z-$axis to change the gradient they
experience. Moving the ions in a linear Paul trap is achieved by
applying suitable voltages to the electrodes that are responsible
for the axial confinement.
Even an always-on interaction (that arises, for instance, when using
permanent magnets to generate a field gradient) used in conjunction
with coherent manipulation of individual spins maybe employed such
that prescribed entangled states can be generated in a time-optimal
way \cite{Fisher2009}.
for the axial confinement. Even an always-on interaction (that
arises, for instance, when using permanent magnets to generate a
field gradient) used in conjunction with coherent manipulation of
individual spins maybe employed such that prescribed entangled
states can be generated in a time-optimal way \cite{Fisher2009}.

\paragraph{Locally variable trap potential}

The advent of micro-structured traps brings with it the possibility
to attain control over local trapping potentials and thus, of
engineering the $J-$coupling locally. For example, employing a
linear Paul trap with dc electrodes that are segmented such that
individual electrodes have an axial extension of the order of the
inter-ion separation allows for shaping the local electrostatic
potential such that long-range coupling is strongly suppressed and
essentially only uniform nearest-neighbour coupling exists
\cite{McHugh2005}.

Furthermore, micro-structured traps with segmented electrodes allow
for detailed local sculpting of $J$-coupling constants. This can be
useful to generate effective time evolution operators making it
possible to reach a desired entangled state in a single time
evolution step without the need for refocussing. Examples how to
generate cluster states in this way are given in
\cite{Wunderlich_H2009}.

A possible drawback of the use of small electrodes is that this
usually implies an equally small distance between ions and nearby
surfaces (i.e., typically of order 10~$\mu$m) which leads to high
heating rates of the ions' secular motion, and thus limits the
useful time available for coherent dynamics \cite{Turchette2000}.
Using ion traps embedded in a cryogenic environment considerably
reduces this heating rate \cite{Deslauriers2006,Labaziewicz2008},
and therefore, may provide a feasible approach to using small
electrode structures. A micro-fabricated linear surface trap cooled
to 4 K that includes current-carrying structures for generating a
magnetic field gradient was recently demonstrated \cite{Wang2008}.
The magnetic gradient was used to address individual $^{88}$Sr$^+$
ions in frequency space on an optical transition between Zeeman
states of the electronic ground state 5S$_{1/2}$ and the metastable
state 4D$_{5/2}$, respectively.

Segmented ion traps also allow for transporting ions between
different trap locations
\cite{Wineland1998B,Kielpinski2002,Rowe2002,Home2006B,Hensinger2006,Schulz2006,Reichle2006,Hucul2008,Huber2008}.
For moving ions, the required electrode structures may be
considerably larger, typically of order 100~$\mu$m, thus helping to
reduce unwanted heating. The transport of ions over small distances
in segmented ion traps gives a further handle to engineer effective
couplings by letting only a subset of ions interact at a time in a
prescribed way. This could be used, for example, to generate
two-dimensional cluster states using a one-dimensional linear
Coulomb crystal \cite{Wunderlich_H2009}.

A possible architecture for a two-dimensional surface trap array
exploiting magnetic field-gradient induced spin-spin coupling as
described above was proposed in \cite{Chiaverini2008}. In
\cite{Clark2008} confinement of $^{88}$Sr$^+$ ions in a  2-d ion
trap array with individual sites spaced apart by about 1.6 mm is
reported. Even though a trap array with such a large inter-ion
spacing does not allow for a useful spin coupling between the ions,
it is nevertheless useful as a proof-of-principle and makes possible
the investigation of the scaling behaviour of this trap array. It
turns out that scaling down of the layered rf lattice trap alone is
unlikely to be a promising route to achieve useful $J-$ coupling. It
appears that further modifications of this trap design are useful to
ensure low enough secular frequencies for sufficiently strong
$J-$coupling (compare equation \ref{eq:J_prop}).

Recently, two-dimensional and linear arrays of Penning traps that
confine individual electrons have been proposed as a promising
physical system for quantum computation and quantum simulations.
Here, the physical origin of the $J-$coupling again is a force that
depends on an electron's spin state and is caused by a magnetic
field gradient and, as in ion traps, the Coulomb force between
electrons mediates the spin-spin coupling
\cite{Ciaramicoli2005,Ciaramicoli2007,Ciaramicoli2008}. Thus,
various couplings as in equation \ref{eq:H_spin_an} with variable
range could be realized with trapped electrons and used, for
instance, to investigate quantum phase transitions.

\subsubsection{Simulating an effective transverse field}
\label{sec:Transverse}

An effective transverse magnetic field can be simulated by driving
transitions between the spin states of ion $i$ employing a radio
frequency or optical field resonant with this ion's qubit transition
\footnote{Optical fields may be used to drive transitions in the
optical range, but also in the RF range by applying a two-photon
Raman excitation}:
 \be
 H_{t} & = & - \hbar \Omega_R \cos(\omega^{(i)}t+\phi') \sigma_x^{(i)} \\
       & = & - \frac{\hbar}{2} \Omega_R \left(\sigma_+^{(i)} + \sigma_-^{(i)} \right)
              \! \left( e^{i\omega^{(i)}t+\phi'} + e^{-i\omega^{(i)}t+\phi'} \right)  \ .
 \label{eq:H_t}
 \ee
Here, $\Omega_R$ is the Rabi frequency that for a linearly polarized
rf field $B_x \cos(\omega_{x}t+\phi')$ reads $\Omega_R= \gamma B_x
/2$ and the resonance condition is $\omega_x=\omega^{(i)}$.

First transforming $H_t$ by $U$, as was done above for the
Hamiltonian $H$ (equation \ref{eq:H_tilde}), then carrying out a
transformation into an interaction picture with respect to
$-\frac{\hbar}{2}\sum_{n=1}^N\omega_n\sigma_z^{(n)} + \hbar
\sum_{n=1}^N \nu_n a_n^{\dagger} a_n$ \cite{Wunderlich2002}, expanding the
resulting Hamiltonian up to first order in $a^\dagger$ and $a$,
neglecting terms that contain $\varepsilon_{in}$ \footnote{In
\cite{Deng2005} the effects of residual spin-phonon coupling caused
by the effective transverse field are calculated}, and applying the
rotating wave approximation, one obtains for spin $i$
 \be
 \tilde{H}_{t}^{(i)} & = & - \frac{\hbar}{2} \Omega_R (\cos\!\phi \, \sigma_x^{(i)}
  - \sin\!\phi \, \sigma_y^{(i)})
 \label{eq:H_t_tilde}
 \ee
when the rf field or optical field is on resonance. Allowing for a
detuning $\omega_{rf} - \omega^{(i)} $ between the applied field and
the spin flip resonance frequency introduces a term proportional to
$(\omega_{rf} - \omega^{(i)}) \sz{}$ on the rhs of equation
\ref{eq:H_t_tilde}. Thus, by driving spin transitions of individual
ions, a variety of additional local effective magnetic fields
$\vec{B_n}' \propto (\cos\!\phi_n,\,-\sin\!\phi_n,\,(\omega_{rf} -
\omega^{(n)})^T$ in Hamiltonian \ref{eq:H_spin_an} may be simulated.
In particular, by choosing the phase $\phi'$ of the rf field such
that $\phi=0$ and fulfilling the resonance condition for all ions, a
global transverse field $B_x'$ for the transverse Ising model
\ref{eq:TI} is implemented.

\subsubsection{Spin-boson model}

The spin-boson model (\cite{Leggett1987,Weiss1999} and references
therein) is the starting point for many theoretical investigations
modeling, for instance, the decoherence of a variety of physical
system. It describes the dynamics of a spin-1/2 coupled to a bath of
harmonic oscillators that represent the spin's environment. Thus,
the spin-boson Hamiltonian contains a term proportional to
$\sigma_z$ (describing the spin's energy), a term proportional to
$\sigma_x$ (describing spin flips), a term describing the harmonic
oscillators, $\hbar \sum_l \nu_l (a_l^{\dagger} a_l +1/2)$, and
finally a term coupling the spin to the oscillators
 \be
 \frac{\hbar}{2} \sigma_z \sum_l C_l Q_l
 \label{eq:S_B}
 \ee
where $Q_l$ is the position coordinate of the $n-$th harmonic
oscillator while $C_l$ describes the coupling strength between this
oscillator and the spin. We note that all these ingredients are
already present in the formulae presented above: Picking out a
particular ion $i$, equation \ref{eq:H_i} may be rewritten as
 \be
 H_I^i &=& - \frac{\hbar}{2} \omega_i \sigma_z^{(i)} -
\frac{\hbar}{2}  \sigma_z^{(i)} \sum_{n=1}^N  \,
\partial_z\omega_i \ q_n \nonumber \\
       &=& - \frac{\hbar}{2} \omega_i \sigma_z^{(i)} -
\frac{\hbar}{2}  \sigma_z^{(i)} \sum_{l=1}^N  \,
\partial_z\omega_i \ \left( \sum_{n=1}^N S_{nl}\right) Q_l
 \label{eq:H_I_i}
 \ee
and we can set $C_l=\partial_z\omega_i \ \left( \sum_{n=1}^N
S_{nl}\right)$ to obtain in equation \ref{eq:S_B} the coupling
between spin $i$ and the harmonic oscillators represented by the
collective vibrational modes of a string of $N$ ions. The term
proportional to $\sigma_x$ is obtained by driving transitions
between the two spin orientations (see below, eq.
\ref{eq:H_t_tilde}).

Now the question arises how to physically single out a particular
spin $i$ from the sum over all spins in the second term of $H_I$
(equation \ref{eq:H_i}) such that we are left with $H_I^i$ in
equation \ref{eq:H_I_i}. This could be done by exerting a
state-dependent force only on ion $i$ as is shown in
\cite{Porras2008}. Porras et al. detail how laser beam(s) focussed
onto one particular ion give rise to optical forces that induce the
required coupling. Furthermore, it is shown how phonon baths
characterized by a variety of spectral densities may be simulated
and how effects caused by the finite size of the ion crystal may be
observed.

\subsection{Phase transition}
\label{sec:Phase}

\subsubsection{Quantum transverse Ising model}

The quantum transverse Ising model (TIM, equation \ref{eq:TI})
displays a phase transition already in one dimension
\cite{Sachdev1999}, even at zero temperature, between an ordered
state and a disordered state upon varying $B_x'/J$. Milburn et al.
\cite{Milburn1999,Milburn2000} proposed an implementation of a
unitary map closely related to the TIM using Raman laser pulses. In
\cite{Porras2004B} it is proposed how to observe this quantum phase
transition (QPT) with trapped ions interacting with a standing laser
wave that creates a long-ranged spin-coupling as outlined above
(section \ref{sec:Spin-spin}). Barjaktarevic et al.
\cite{Barjaktarevic2005} proposed to implement a unitary map that
displays a QPT belonging to the same universality class as the
transverse Ising model.

In order to get a qualitative picture of such a phase transition we
first consider the collection of spins described by $H_{TI}$
(equation \ref{eq:TI}) in the limit where the effective transverse
field, $B_x'$ assumes a finite value, but the mutual coupling $J$
between spins vanishes, $B_x' \neq 0 = J$, such that, at zero
temperature, all spins are oriented along the $x-$direction and the
groundstate $\ket{0} = \prod_i\ket{\rightarrow}_i=
\prod_i(\ket{\uparrow}_i + \ket{\downarrow}_i)$ of the system is a
product of eigenstates of $\sigma_i^x$. Consequently,
$\bra{0}\sigma_i^z \sigma_j^z \ket{0}=\delta_{ij}$, that is, the
orientation of different spins with respect to the the $z-$axis is
uncorrelated. On the other hand, if the spin-spin coupling doesn't
vanish, $B_x'\gg |J| >0$, the spins will acquire a tendency to align
along the $z-$axis. Thus, the ground state of the system will start
to exhibit non-vanishing, but short ranged correlations between
spins located at sites $r_i$ and $r_j$, respectively, with a
characteristic length scale $\xi$ \cite{Sachdev1999}
 \be
 \bra{0}\sigma_i^z \sigma_j^z \ket{0} \propto  e^{-|r_i - r_j|/\xi} \ .
 \label{eq:Corr_Bx}
 \ee
Here, $\ket{0}$ is the system's ground state when $B_x' \gg |J| \neq
0$.

Now, we turn to the limit $B_x' \ll J > 0$. If $B_x'$ vanishes
completely, then an infinite system of spins possesses two
degenerate ferromagnetic ground states where all spins are either
parallel or anti-parallel to the $z-$axis, and the total
magnetization of the system is $M_0$. Turning on a small field $B_x'
\ll J$ pointing in the $x-$direction will tend to reorient some of
the spins along the $x-$direction (classically, their magnetic
moments will start to precess around the $x-$direction, thus
flipping spins between the two possible $z-$orientations). The
ground state of the system, $\ket{0}$  is still close to the
$B_x'=0$ case and is obtained from a perturbative expansion in the
small parameter $B_x'/J$. One finds $\bra{0}\sigma_i^z \ket{0}=
M/M_0$ where $M<M_0$ is now the spontaneous magnetization of the
system. Again calculating the correlation of the orientation of
different spins in the $z-$direction for this ground state gives
\cite{Sachdev1999}
 \be
  \lim_{|r_i - r_j|\rightarrow \infty} \bra{0}\sigma_i^z \sigma_j^z \ket{0}
   = \left(\frac{M}{M_0}\right)^2 \ .
 \label{eq:Corr_J}
 \ee

The expressions \ref{eq:Corr_Bx} and \ref{eq:Corr_J} for the spin
correlation are incompatible in the sense that a continuous
variation of $B_x'/J$ between the two limiting cases considered
above doesn't allow for an analytical connection between the two
expressions on the right-hand-sides of equations \ref{eq:Corr_Bx}
and \ref{eq:Corr_J}. Thus, there must exist a critical value of
$B_x'/J$ where the correlation changes between these two
expressions. This is the case even at zero temperature being
indicative for a quantum phase transition.

An experimental realization of the proposal by Porras and Cirac
\cite{Porras2004B} with two trapped and laser cooled $^{25}$Mg$^+$
ions was recently reported by Friedenauer et al.
\cite{Friedenauer2008}. Here, the coupling $J$ between spins is
induced by a state-dependent ac Stark shift caused by a "walking"
standing wave. Two hyperfine states of the electronic ground state
of each $^{25}$Mg$^+$ ion served as an effective spin-1/2 with
eigenstates $\ket{\downarrow}$ and $\ket{\uparrow}$. In brief, the
experimental procedure was as follows: Optical pumping is employed
to first initialize the two ions in state
$\ket{\downarrow,\downarrow}$. Resonantly driving the transition
between the two hyperfine states with RF radiation simulates an
effective transverse field $B_x'$ (section \ref{sec:Transverse}) and
allows for preparation of the initial paramagnetic state
$\ket{\rightarrow, \rightarrow}$ (both spins are oriented along the
effective $B_x'$ field). Then, the $J-$coupling  between the two
spins is ramped up adiabatically from zero to a desired value such
that the ratio $J/B_x' \leq 5.2$. For $J \gg B_x'$ one expects that
the two spins are found in a ferromagnetic state, that is, aligned
along the $z-$axis. In general, the new ground state will be a
superposition of the states  $\ket{\downarrow,\downarrow}$ and
$\ket{\uparrow,\uparrow}$ which is indeed found experimentally. In a
further experiment the two spins were prepared initially
anti-parallel to the effective $B_x'$ field, an excited paramagnetic
state. Turning on the spin-spin coupling adiabatically then resulted
in an excited state in the presence of strong $J-$coupling ($J/B_x'
\gg 1$): the anti-ferromagnetic state $\ket{\uparrow,\downarrow}
+\ket{\downarrow,\uparrow}$. In both cases the entanglement of the
final state was proven experimentally.

\subsubsection{Scalar fields}

Classical phase transitions between spatial configurations of ions
confined in a linear or ring shaped trap have been investigated
experimentally and theoretically
\cite{Waki1992,Raizen1992,Birkl1992,Dubin1993,Schiffer1993,Drewsen1998,Kjaergaard2003,Morigi2004,Morigi2004B,Fishman2008}.
Retzker et al. \cite{Retzker2008} suggest to explore a phase
transition between a linear and a zig-zag configuration
\cite{Fishman2008} in the quantum regime, that is, with the ion
string cooled close to its vibrational ground state. The ion trap
parameters are chosen such that a linear Coulomb crystal forms
(i.e., the confinement in the axial direction is weaker than in the
radial direction), and, in addition, $\nu_{y,1} \gg \nu_{x,1}$ such
that one radial degree of freedom is essentially frozen out after
laser cooling, but the other may be employed for quantum
simulations. Linear Paul traps as well as Penning traps where the
ions are confined on the axis of rotational symmetry
\cite{Powell2002} are well suited for this purpose.

For $\nu_{1,x}$ below a critical value $\nu_{1,x}^c$, a zig-zag
configuration of the ions will form with two degenerate states. It
is shown in \cite{Retzker2008} that coherent tunneling between these
two states maybe observed and used for interferometry. Furthermore,
by adjusting $\nu_{1,x}$ around $\nu_{1,x}^c$ the parameters of a
Hamiltonian describing the ions' motional excitation along the
$x-$direction may be determined. This Hamiltonian describes at the
same time a non-linear scalar field and by adjusting $\nu_{1,x}$ the
effective mass of this field may be varied, in particular in the
vicinity of a phase transition. An attractive feature of this
proposal is that {\it global} adjustment of the radial trapping
frequency $\nu_{1,x}$ is sufficient and no local control over the
trapping potential is needed.

\subsection{Neural Network}
In \cite{Pons2007,Braungardt2007} it was shown that a linear Coulomb
crystal of laser cooled trapped ions may be used to implement a
neural network (NN) -- according to the Hopfield model
\cite{Hopfield1982} -- allowing for robust storage of classical
information, and furthermore, that such a NN may be employed for
error-resistant quantum computation. We will first briefly introduce
some features of the Hopfield model (see, e.g., \cite{Rojas1996}
chapter 13 and references therein) and then outline how some
features may be realized with trapped ions.

\begin{figure}[tbh]
    \centering
        \includegraphics[width=\columnwidth]{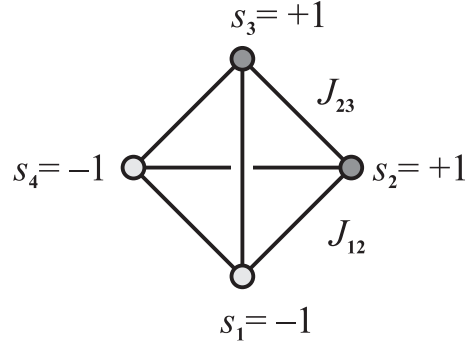}
    \caption{Illustration of a neural network that consists of four nodes according
    to Hopfield \protect\cite{Hopfield1982}. Each node
    may exist in one out of two states indicated by $s_i=\pm1$,
    $i=1,2,3,4$. The nodes are all interconnected with $J_{ij}=J_{ji}$.
    Here, the network forms a pattern $\vec{\xi}= (-1,+1,+1,-1)^T$.
    (Note that the number $p$ of patterns that can be stored by a Hopfield network
    is about $p=0.14 N$ where $N$ is the number of nodes).
    }
    \label{fig:tetraeder}
\end{figure}

A neural network according to the Hopfield model consists of $N$
nodes (corresponding to neurons) that are all mutually
interconnected and we indicate the (synaptic) strength of the
connection between node $i$ and node $j$ by $J_{ij}$. This
connection is symmetric, that is $J_{ij}=J_{ji}$ and the nodes do
not feed back to themselves, $J_{ii}=0$. Below we will see what
exactly the meaning of the coupling strengths $J_{ij}$ is. Each node
can assume one of two possible states which we label by $s_i=\pm1$
(compare figure \ref{fig:tetraeder}). When modeling neurons, this
would indicate whether the neuron is "firing" (a sudden change of a
biologically relevant potential difference) or in a quiescent state
at a given time. Furthermore each node is characterized by a
threshold $h_i$ that, in addition to the couplings $J_{ij}$,
influences its response to the the input from other nodes.

A characteristic feature of a Hopfield NN is that it can store
information distributed over all nodes. Thus, it is robust against
errors in individual nodes or even the loss of some nodes and
against thermal noise. The dynamics of the NN, being determined by
the strengths of interactions between all nodes, brings the system
back to one out of a selection of particular states (that we call
patterns) once this state was disturbed. This feature may be taken
advantage of for pattern recognition: When the NN is presented with
some input state, which is done by initializing all nodes with a
given value, then the NN evolves into the pattern that is was
trained to recognize that is closest to this initial state. So how
does the NN dynamically evolve?

We start at time $t$ with an initial global state of the NN labeled
by the $N-$component vector $\vec{s}$ where each local node is
characterized by $s_i$. Then the state of each node at time
$t+\delta\!t$ is determined by the sign of the activation function
$A$ at time $t+\delta\!t$:
 \be
 s_i(t+\delta\!t) = \rm{sgn}(A(t+\delta\!t)) \ ,
 \label{eq:s_delta_t}
 \ee
where
 \be
 A(t+\delta\!t)= \sum_{j\neq i} J_{ij}s_j(t) + h_i \ .
 \label{eq:A}
 \ee
This means that at time $t+\delta\!t$, node $i$ will be firing
($s_i=+1$) or in a quiescent state ($s_i=-1$) depending on the state
of all other nodes at time $t$, weighted with the strength $J_{ij}$,
and the individual threshold $h_i$. One of the two following
alternative model assumptions is usually made: The updating of node
$i$ may either occur sequentially, one node after the other, or in
parallel all at the same time.

We now define the cost function
 \be
 H_H(\vec{s}) = -\frac{1}{2}\sum_{i,j (i\neq j)} J_{ij}s_i s_j -
                \sum _i h_i s_i
 \label{eq:H_H}
 \ee
and it can be shown \cite{Rojas1996} that an evolution of the
network according to the dynamical rule for sequential updating
\ref{eq:s_delta_t} (together with the definition \ref{eq:A}) leads to
a monotonous decrease of $H_H$ \footnote{In this brief account we
consider the system at zero temperature only}. This means that
flipping spins (in agreement with the rules) drives the system
towards local or global minima in the "landscape" defined by $H_H$
over the space of $\vec{s}$.

Immediately we realize that the expression defining $H_H$ is
formally equivalent to the spin Hamiltonian $H_s$ (equation
\ref{eq:H_spin_an}) considering the case of spin-1/2 particles with
$J_{ij}^{(x)}= 0 = J_{ij}^{(y)}$. As noted at the beginning of the
section on spin models, further simplifications of $H_s$, namely
uniform $J-$coupling, only nearest-neighbour interaction, and no
transverse field lead to the Ising model (equation \ref{eq:Ising}).
When considering the Ising model one usually assumes that all
coupling constants are positive or all are negative and thus arrives
at a limited set of minima (e.g., for $J>0$ all spins are all either
up or down). However, here the couplings $J_{ij}$ are not restricted
to being all positive or all negative, thus allowing for a rich
"energy landscape" with many minima.

Our goal is to "teach" the NN a specific pattern,  $\vec{\xi}$,
$\xi_i = \pm 1$ towards which the NN is supposed to relax, if it is
displaced from this pattern, or if presented with a similar input
pattern. In other words, this pattern must correspond to a minimum
of $H_H$. Where exactly this minimum occurs, that is, for which
particular configuration $\vec{s}$, is determined by the choice of
the interaction strengths $J_{ij}$. Therefore, if one wants the NN
to "remember" a particular pattern $\vec{\xi}$, (i.e., create a
minimum of $H_H$ for a specific arrangement $\vec{\xi}$ of the
nodes), the $J-$couplings need to be adjusted accordingly. At first
sight (and probably still at second) it is not obvious how a set of
$J-$couplings may be found that creates a minimum of $H_H$ for a
desired pattern and thus makes the NN evolve towards this pattern.
It turns out that using Hebb's rule \cite{Hebb1949,Mezard1987} to
determine the values of the couplings $J_{ij}$ gives just the right
set of $J-$couplings.

In order to memorize a pattern $\vec{\xi}$, each $J_{ij}$ is
modified according to Hebb's rule of learning, such that
 \be
 J_{ij}^{new} = \lambda J_{ij}^{old} + \epsilon \xi_i\xi_j \ .
 \label{eq:J_Hebb}
 \ee
When starting with an "empty" memory, then $J_{ij}^{old} = 0$. In
the original Hopfield model $\lambda=1=\epsilon$; another choice is
$\epsilon=1/N$. With the set of interconnections, $J_{ij}$
determined by equation \ref{eq:J_Hebb} the NN, upon turning on its
dynamics, will eventually end up in the memorized pattern
$\vec{\xi}$. If the NN is supposed to memorize $p$ patterns,
$\vec{\xi}^\mu$ with $\mu=1,2,3,\ldots,p$ instead of just a single
pattern, then the learning rule is modified, namely
 \be
 J_{ij} =  \epsilon \sum_{\mu=1}^p \xi_i^\mu \xi_j^\mu \ .
 \label{eq:J_Hebb_sum}
 \ee
Here, we start with an empty memory, i.e. initially $J_{ij}= 0
\forall i,j$. After fixing the strength of the pairwise $J-$coupling
in this way, the patterns are "memorized" by the NN, and, when
presented with some initial state, the NN will evolve towards a
minimum of $H_H$ associated with one of these patterns. The upper
limit for the number of patterns that such a network may faithfully
recall via its associative memory is about $0.14N$ \cite{Rojas1996}.

Now, it will be discussed how a linear Coulomb crystal of trapped
ions may be employed to realize a Hopfield-type NN. We have already
noted the similarity between $H_H$ (equation \ref{eq:H_H}) on one
hand and the two spin Hamiltonians $H_s^{(a)}$ (equation
\ref{eq:H_spin_an}) and $H_s^{(ion)}$ (equation \ref{eq:H_s_ion})
for $J_{ij}^{(x)}= 0 = J_{ij}^{(y)}$ on the other. A further
similarity can be seen between equation \ref{eq:J_Hebb_sum},
defining the synaptic strengths in the Hopfield model and equation
\ref{eq:J} where the ion spin coupling constants are given. We
rewrite equation \ref{eq:J} using the expression \ref{eq:eps}
assuming a constant field gradient and obtain
 \be
 J_{ij}^{(ion)} = \frac{\hbar}{2m} \partial_z\omega
                  \sum_{n=1}^N \frac{1}{\nu_n^2} S_{in} S_{jn}
 \label{eq:J_ij_ion}
 \ee
where the matrix elements $S_{in}$ may assume positive and negative
values, however usually different from $\pm 1$ as in the generic
Hopfield model. Also, the sum in equation \ref{eq:J_ij_ion} extends
over all $N$ vibrational modes instead of $p$ patterns, and the
contribution to $J_{ij}^{(ion)}$ from each vibrational mode $n$ is
weighted by the square of the respective vibrational frequency
$\nu_n$. Despite these differences, it is possible to store spin
patterns with trapped ions in a robust way.

What are the patterns, based on the learning rule \ref{eq:J_ij_ion},
that can be stored by a trapped ion neural network? The orientation
of a specific ionic spin $i$ in a pattern $\vec{\xi}$ of spin
orientations is determined by the sign of $S_{in}$ for a given
vibrational mode $n$. In case of the COM mode, for example, we have
$S_{in}> 0$ (or $S_{in}< 0)$ $\forall i$ which means that in the two
associated patterns all spins are either pointing up or all down.

Numerical simulations \cite{Pons2007,Braungardt2007} show that, in
order to robustly store a pattern of spin orientations it is
necessary to adjust the trapping potential such that the vibrational
frequencies $\nu_1$ and $\nu_2$ (that appear in the learning rule
\ref{eq:J_ij_ion}) are nearly degenerate. This can be achieved by
using ion traps with segmented electrodes where the Hessian of the
trapping potential can be tuned by applying suitable voltages to
individual segments \cite{McHugh2005,Wunderlich_H2009}. The
robustness of the ion NN has been quantified by Monte Carlo
simulations for a string of 40 $^{40}$Ca$^+$ ions trapped such that
the two lowest vibrational modes characterized by $\nu_1$ and
$\nu_2$ are nearly degenerate. For this purpose, a spin pattern
$\vec{\xi^\mu}$, $\mu=1,2$ corresponding to either one of the two
modes is prepared as the initial condition of the ions' spin
orientations. Then a fixed number $r$ of spins are randomly selected
and their state is inverted. This is followed by the time evolution
of the ion string until an equilibrium state is reached. The
probability of the ion string for ending up in the initial pattern
is found to be more than 97 \% even when up to eight randomly
selected spins are initially in the wrong state (i.e., they are not
matching the desired pattern). Thus, even with 20\% of the spins
being flipped, the ion string returns towards a desired pattern.

So far we have looked at the robustness of a Hopfield-type neural
network where each node consists of a single ion and which is used
for classical information storage. For a {\em quantum} neural
network, the elementary unit for information processing is, instead
of a single spin-1/2 with two eigenstates corresponding to $\ket{0}$
and $\ket{1}$, a pattern $\vec{xi}$ of $N$ spins. In
\cite{Pons2007,Braungardt2007} it is detailed how  an appropriately
chosen set of patterns with eight trapped ions may be used as qubits
thus forming a quantum neural network. In particular, single-qubit
gates and two-qubit gates by adiabatic passage are worked out that
make use of additional effective fields.

\subsection{Mesoscopic systems}

The question if and to what extent quantum mechanical effects play a
role in mesoscopic or even macroscopic systems was posed already
shortly after quantum theory was developed \cite{Schroedinger1935}.
Except concerning some particular physical phenomena, for example,
macroscopic currents in a superconducting material, it is usually
assumed that superposition states, interference, and entanglement
are not relevant when dealing with entities made up from a large
number of individual constituents that, in addition, may interact
with a noisy environment at non-zero temperature, for example,
biologically relevant molecules. Indications that quantum mechanics
indeed plays an important role even for large molecules come from
recent experiments where interference with C$_{70}$ molecules was
observed \cite{Hackermuller2004}.

A theoretical study of how dephasing caused by coupling to a noisy
environment can actually assist in transferring energy (or
information), instead of impeding it, across a network of quantum
nodes is presented in \cite{Plenio2008}. It is shown that, in a
simplified model, light-harvesting molecules may take advantage of
this process. An experimental verification directly with these
molecules is difficult at present. However, trapped ions may be used
to implement a suitable model to test such a prediction. For a first
simple demonstration, appropriately chosen internal levels of a
single ion may be used to model the nodes of a network and induced
and spontaneous transitions between these levels to mimic energy
transport.

Another recent theoretical study demonstrates that entanglement may
persist in a model two-spin molecule even when it is exposed to a
decohering hot environment, a situation typically encountered for
biological molecules \cite{Cai2008}. The time dependent Hamiltonian
modeling the two-spin molecule used for this study is
 \be
 H_M(t) = J(t) \sigma_x^{1} \sigma_x^{2} + B(t)(\sigma_2^{1} +
 \sigma_z^{2}) \ .
 \ee
This molecule is then coupled to a noisy environment modeled by a
master equation. After performing a unitary transformation such that
$\sigma_x \rightarrow \sigma_z$ and $\sigma_z \rightarrow -\sigma_x$
we recognize a time-dependent variant of the quantum transverse
Ising model \ref{eq:TI} which could be realized with an ion spin
molecule by using the spin-coupling present in equation
\ref{eq:H_tilde} and by adding a driving field as in equation
\ref{eq:H_t_tilde}.

\section{Quantum Optics Simulating Quantum Optics}
\label{sec:SimulatingOptics}

The formal similarities for creation and annihilation operators for photons and phonons on one side, and spin $S=1/2$ particles on the other side suggests the equivalence of various and seemingly different experiments within the field of quantum optics. In this notion, a Ramsey measurement might be regarded as simulating a Mach-Zehnder interferometer, since a wavefunction is coherently split and recombined after a independent time-evolution of its constituents. Therefore one general proposition which holds also for other classes of experiments discussed in this paper is maybe of particular importance here: simulation goes both ways. If the Hamiltonian describing a setup in an ion trap, possibly including static and radiation fields, is equivalent to the Hamiltonian of another system, then the ion trap experiment can be thought as a simulation of the other system of interest. But on the other hand it also means, that these other system might reveal some insights about certain aspects and experiments in ion traps, and, depending on the specific technical difficulties when trying to perform it in an ion trap, it might also be imaginable to simulate some aspects of ion trap physics in other systems. As pointed out earlier, however, trapped laser-cooled ions are a flexible, well controlled and well understood system, which can be brought into a variety of regimes, where the physics resembles other systems. Nowadays we are familiar with the fact that both photons but also the internal states of particles can be entangled or combinations of them and it might sound unusual to phrase any of such experiments as a simulation of the other. Nevertheless the creation and detection of entanglement of particles or phonons is subject to different problems and certain aspects of entanglement and its meaning in investigating fundamental questions of quantum mechanics might best be looked at in either approach \cite{Wineland1998A}.

\subsection{Theoretical Background}
\label{sec:SimOptBackground}

Already in the early days of quantum information it became clear that quantum computing at a level of factorizing large numbers would be a \textit{daunting task} (see \cite{Wineland1998A}), though, since then, astonishing progress was made this field. The first articles considering quantum simulations were, to some extent, motivated by the insight that the requirements for quantum simulations (in terms of gate fidelity etc.) are less stringent despite their potential to investigate systems that are beyond reach of simulations with classical computers. Despite all progress, this statement still holds and is one motivation for this topical review.


The experiments in quantum optics that are considered in \cite{Wineland1998B,Leibfried1997} can be simulated by trapped ions interacting with radiation fields that drive Raman transitions between hyperfine levels. A similar description can be obtained when this transitions are directly driven using microwaves (using auxiliary inhomogeneous magnetic fields) \cite{Mintert2001,Mintert2001E,Wunderlich2002,Johanning2009} or when optical transitions between metastable states are used. For a harmonically bound atom interacting with two travelling wave light fields, both detuned from any real state and with frequency difference $\Delta\omega$ and phase difference $\phi$ the Hamiltonian in the interaction picture can be brought into a form $H_{\epsilon l}$ (after rotating wave approximation and adiabatic elimination of the near resonant excited states and in the Lamb-Dicke limit $\eta^2\left\langle(a+a^\dagger )\right\rangle \ll 1$)  \cite{Wineland1998B}
with integer $l$ indicating changes in the motional state, and $\epsilon \in[0,1]$ indicating changes in the internal state. This Hamiltonians are frequently used in experiments with laser cooled ions; to name a few, $H_{10}$ is called the carrier transition, $H_{12}$ is the second blue sideband, $H_{01}$ is called coherent drive, $H_{02}$ is the squeeze drive. Nesting and concatenation operators of the set $H_{01}, H_{02}, H_{03}, H_{10}, H_{11}, H_{12}, H_{13}$ is sufficient to efficiently generate a wide range of Hamiltonians, as, for example the Hamiltonian for a spin $s=1/2$ particle with mass $\mu$ in an arbitrary potential \cite{Leibfried2002}.

These chaining of Hamiltonians could be used for a variety of experiments, as, just to highlight a few, the phonon maser \cite{Wallentowitz1996} and three phonon down conversion \cite{Wineland1998A} and has been used to create motional cat staes \cite{Monroe1996B}, coherent states \cite{Meekhof1996A,Meekhof1996B}, and to create a Hamiltonian similar to two photon excitation in cavity QED \cite{Leibfried1997}. An example discussed in the following subsection is an an experimental analogy to a nonlinear two beam (e. g. Mach-Zehnder) interferometer as proposed in \cite{Wineland1998A}. Using ions for the implementations allows for high state preparation and detection efficiency and eliminates the need for data postselection.

All the experiments above used only a single ion and innumerable possibilities come into reach using multiple ions. The first demonstration of the CNOT gate \cite{Cirac1995} was a step in this direction using, however, a single ion and its motional degree of freedom \cite{Monroe1995}. This concept can be adapted to entangle different atoms when it is carried out after the state of one ion is mapped onto the centre of mass (CM) mode and afterwards mapped back \cite{Schmidt-Kaler2003,Leibfried2003B}.

\subsection{Experiments}
\label{sec:SimOptExperiments}

\begin{figure}[t!]
    \centering
       \includegraphics[width=\columnwidth]{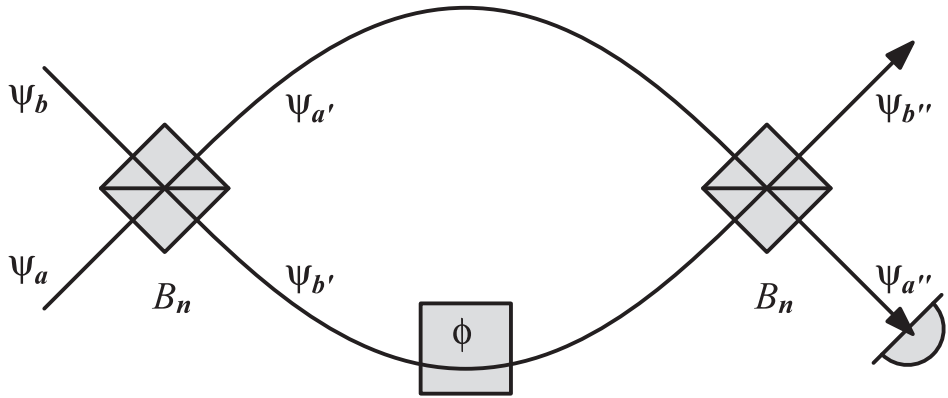}
    \caption{Ion trap Mach-Zehnder interferometer: the manipulation of the
wavefunction of a trapped ion can be thought analogous to a set of
two nonlinear beamsplitters forming a nonlinear Mach-Zehnder
interferometer. }
    \label{fig:machzehnder}
\end{figure}

As a demonstration of a quantum simulation with a single ion, the Hamiltonians $H_{11}, H_{12}, H_{13}$ are used in \cite{Leibfried2002} to simulate a set of two nonlinear beam splitters forming a nonlinear Mach-Zehnder.  A single $^9 Be^+$ ion provides the internal state $\ket{a}$ with
\begin{equation}
\begin{array}{rcl}
\ket{0}_a & = & \ket{F=1, m_F =-1}\\[2mm]
\ket{1}_a & = & \ket{F=2, m_F = -2}
\end{array}
\label{eq:MZinternalStates}
\end{equation}
and the motional state $\ket{n}_b$ labelling the $n$th vibrational state. The experimental sequence is depicted in \ref{fig:machzehnder} and is as follows: the ion is cooled to the ground state of motion $\ket{1}_a \ket{0}_b$. A Raman $\pi/2$ pulse applied on the appropriate sideband is used to create the state proportional to $\ket{1}_{a'} \ket{0}_{b'} + \ket{0}_{a'} \ket{n}_{b'}$. This is equivalent to a beamsplitter: analogous to coherently splitting light into two separate directions, the population is coherently split into two states which, in this case, differ in both internal and motional quantum number. Subsequently the potential of the endcap electrodes is changed yielding trap frequency change $\Delta\nu_{z,1}$ for a time $t$ which introduces a phase shift $n\Delta\nu_{z,1}t=n\phi$ when the ion is in the $n$th vibrational level. Finally a second $\pi/2$ pulse on the same sideband is applied which maps the phase difference to the population in the internal states $\ket{0}_a$ and $\ket{1}_a$ which is measured by Doppler cooling and fluorescence detection. The absolute acquired phases are irrelevant as is the total arm length of a Mach-Zehnder interferometer: longer arms only tend to make it more susceptible to perturbations, however, the signal at the outputs is given by the differential phase shift. The sequence is repeated to improve signal to noise ratio and to extract the sinusoidal time evolution in the different states.

The experiment can be interpreted as nonlinear Mach-Zehnder interferometer. A nonlinear beamsplitter can annihilate a photon in mode $a$ while creating $n$ photons in mode $b$ (or vice versa). The annihilation operator $a$ is replaced in this experiment by the atomic raising operator $\sigma^+$ between the states $\ket{F=2, m_F = -2}$ and $\ket{F=1, m_F =-1}$. These two operators are not equivalent but they act the same as long as the $\{\ket{0}_a,\ket{1}_a\}$ subspace is not left, which is fulfilled as long as the input state is $\ket{1}_a \ket{0}_b$. The optical mode with lowering operator $b$ is replaced by the motion along a selected direction and phonon number states $\ket{n}_b$. A $n$th order nonlinear beamsplitter is then given as
\begin{equation}
    B_n = \hbar\Omega_n \left[a\left(b^\dagger\right)^n + a^\dagger\left(b\right)^n\right].
    \label{eq:MZnonlinearBeamsplitter}
\end{equation}
When this operator brackets an operator that causes a phase shift in one arm by $n\phi$, the system resembles a Mach-Zehnder interferometer and the probability of detecting a phonon in output $a''$ becomes
\begin{equation}
    P_{a''} = \frac{1}{2}\left[1-\cos\left(n\phi\right)\right]
    \label{eq:MZfringes}
\end{equation}
This experiment requires a sideband transition in the $n$th sideband, which can be very inefficient with increasing $n$ as the sideband Rabi frequencies scale with $\eta^n$ and the preparation becomes slow. An faster alternative route exploiting an auxiliary state and first order sidebands is detailed in \cite{Wineland1998A}.

An increase in fringe frequency proportional to the order of the nonlinearity is found but the fringe contrast is reduced with increasing $n$, as preparation and manipulation imperfections spoil the fidelity of the states and the detection efficiency is below one for all experiments. The sensitivity of the output signal to phase changes is maximized at the steepest slope of the signal and the signal to noise ratio for the $n=2$ and $n=3$ measurements is found to be below a perfect first order interferometer with unit detection efficiency. A potential improvement by increasing $n$ from 2 to 3 is almost perfectly cancelled by the reduction of fringe contrast in the data presented in \cite{Leibfried2002}.

\section{Transverse standing wave: Bose-Hubbard-physics (and beyond)}
\label{sec:BHM}

Laser cooling and trapping has been considered \cite{Haensch1975,Wineland1975} and successfully applied (see \cite{Ghosh1995,Metcalf2001}, and references therein) to both neutral atoms and ions. Both types of laser cooling experiments are often very much alike in terms of the description of the experiment (trap frequencies and depths, Rabi frequencies etc.) but also in terms of experimental techniques. A recent topic in cold neutral atom physics is the combination of cold trapped atoms and optical lattices which allows to experimentally investigate the Bose Hubbard model \cite{Fisher1989}, which is related to the Hubbard model in solid state physics \cite{Hubbard1963}.  The freedom in changing tunnelling and interaction parameters, the defect free optical lattices and the controlled introduction of disorder \cite{Schulte2006} allow nice and clean experiments \cite{Greiner2002,Morsch2006}. Also fermionic atoms have been loaded into traps and lattices and brought to quantum degeneracy. The description of such experiments by the Bardeen-Cooper-Schrieffer (BCS) theory \cite{Bardeen1957,Giorgini2008} is known from superconductivity and this analogy encourages the hope that this experiments can help to investigate and understand high $T_c$ superconductivity \cite{Hofstetter2002,Holland2001}.

The large and long range Coulomb force being the dominant force in ion traps has a number of consequences: trap frequencies, trap depths, interaction energies, and particle separations are typically higher by orders of magnitude compared to traps for neutral atoms. Thus Bose-Einstein condensation (and also Bose-Hubbard physics), one long sought holy grail in quantum optics, has therefore been investigated with neutral atoms but not with ions: when cooling neutral atoms at sufficient density, their wave functions can overlap and a macroscopic population of the trap ground state forms \cite{Anderson1995,Davis1995B}. When on the other hand an ion chain is cooled to the lowest vibrational mode \cite{Diedrich1989}, the ion separation is orders of magnitude larger than the spatial extension of the wave function.

But quantum degeneracy and Bose Hubbard physics can still be simulated in ion traps when the ions take the role of lattice sites which can be populated by phonons, more specifically excitations of the vibrational modes transverse to the weak trap axis. The phonons therefore represent the neutral atoms located in one lattice site and phonons can tunnel between sites as do the atoms in optical lattices. Furthermore, as we will see later on, the phonons can be made interact by anharmonicities in trapping potential. This anharmonicities can be introduced by the light shift of an off-resonant standing wave. This could be used to demonstrate a Bose-Einstein condensate (BEC) of phonons \cite{Porras2004A}, the superfluid-Mott insulator transition \cite{Deng2008} and, at higher levels of interaction energy, the Tonks-Girardeau gas \cite{Deng2008} and even frustrated $XY$-models \cite{Schmied2008}. In 1D, BEC at non-zero temperature is only possible when the number of atoms (or here phonons) is finite, and we are in the weak coupling regime, which means, that the average interaction energy is smaller than the expectation value of the kinetic energy. A peculiarity of the effective phonon interaction being dependent on the position of the standing wave is that it can be tuned, even in sign, similar as a the magnetic field can change the scattering length near a Feshbach resonance.

Besides the general appeal to demonstrate Bose-Einstein condensation in yet another system, Bose-Hubbard physics in ion traps offers advantages, and also drawbacks, over present neutral atom experiments: experiments in ion traps do not only offer global control of tunnelling rate and interaction energy (as we will see later), equivalent to ramping the optical lattice depth or applying a magnetic field to obtain a certain scattering length near a Feshbach resonance. Ion traps allow also for local control: irregular lattices correspond to site dependent tunnelling rates and can be obtained in segmented microtraps; the interaction energy can as well be made site dependent by having independent standing wave intensities and phases at the different ion positions -- in terms of neutral atoms this would correspond to arbitrary 3d Feshbach magnetic fields. Other interesting features include single site control and readout of the state and the potential to extract the density matrix either partially or, even fully (in small systems). On the other hand the experiments discussed here are restricted to bosonic phonons  and thus Bose-Hubbard physics and today's ion traps are mostly restricted to one dimension but true two-dimensional systems can be obtained in surface traps \cite{Seidelin2006}.

\subsection{Background}
\label{sec:BHMBackground}

When we speak of phonons in the context of linear ion traps, we often think in terms of axial normal modes. In this situation momentum is transferred by local interaction with the addressed ion which couples efficiently to the motion of the whole chain due to changing separations and acting Coulomb forces. In this chapter radial modes of the ion motion are considered. Classically a radial displacement of a single ion excites both radial motion of other chain members but also the axial motion of the whole chain. But when the axial ion separations are much larger than radial excursion of the ion motion, the coupling of this motion to all other modes is very inefficient: the motion is almost perfectly perpendicular to the Coulomb interaction between ions all the time and does not change the ion separations to first order. Thus radial phonons can be viewed as being located to one ion but the non vanishing coupling makes the localization non permanent and re-emerges as a finite phonon tunnelling rate.

\begin{figure}[t!]
    \centering
        \includegraphics[width=1.0 \columnwidth]{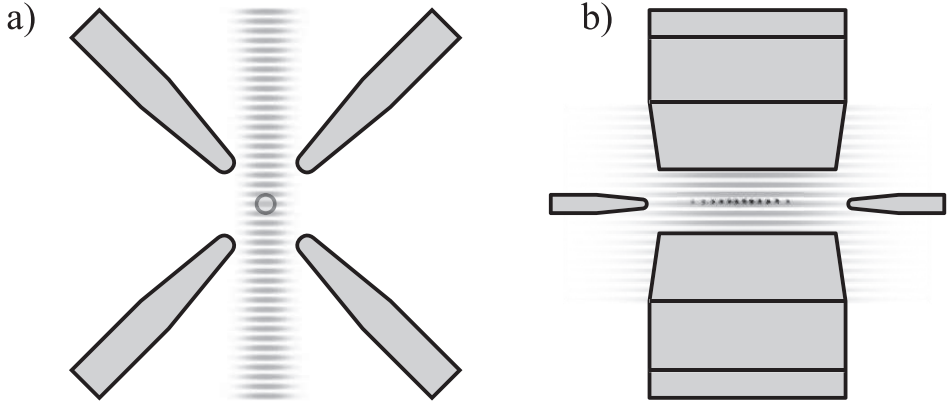}
    \caption{Experimental setup (not to scale) for the simulation of Bose-Hubbard physics. Images a) and b) are sketches of the trap and the applied standing wave potential. The illustrations are not drawn to scale, especially the wavelength of the standing wave is greatly exaggerated for visual clarity. Figure a) shows the trap as seen from a one of the endcap electrodes. Figure b) shows the setup from the side, as seen in most cases from an imaging detector; the weak trap axis is perpendicular to the standing wave.}
    \label{fig:BHMsetup}
\end{figure}

The Hamiltonian describing the system reads \cite{Deng2008}
\begin{equation}
H_0 = \sum_{i=1}^{N} \frac{p_i^2}{2 m} + V_T + \sum_{i,j=1, i>j}^{N} \frac{1}{4\pi\epsilon_0 } \frac{e^2}{|\vec{R}_i - \vec{R}_j |}
\label{eq:ion_chain_Hamiltonian}
\end{equation}
where the trapping potential $V_T$ could be simply quadratic like the effective potential created by a simple linear Paul trap, or be array of microtraps achieved by micro structuring \cite{Rowe2002,Stick2006}. We will look in more detail into one radial coordinate $x$ and excitations at the trap frequency $\nu_{x,1}$ in this direction. The trap frequency $\nu_{y,1}$ can be chosen to be non-degenerate to allow for addressing specific sidebands and to avoid unwanted couplings. The absolute value of the Coulomb force between two particles separated by a large distance $\delta z$ in $z$-direction and displaced only slightly by $\delta x$ in the $x$-direction is
\begin{equation}
|\vec{F}| = \frac{e^2}{4\pi\epsilon_0} \frac{1}{\delta x^2 + \delta z^2} \approx \frac{e^2}{4\pi\epsilon_0} \frac{1}{|\delta z|^2}
\label{eq:CoulombForce}
\end{equation}
The component in $x$-direction of the Coulomb force is given by
\begin{equation}
F_x = |\vec{F}| \frac{\delta x}{\sqrt{\delta x^2 + \delta z^2}} \approx |\vec{F}|\frac{\delta x}{|\delta z|}.
\label{eq:BHMforceRadial}
\end{equation}
When the radial separation $\delta x$ is identified with $x_i - x_j$ and the axial separation $\delta z$ with the unperturbed $z_i^0 - z_j^0$, the radial potential for small relative displacements becomes
\begin{equation}
H_{\textit{Coulomb,x}} \approx - \frac{1}{2} \frac{e^2}{4\pi\epsilon_0} \sum_{i,j=1, i>j}^{N} \frac{\left(x_i - x_j\right)^2}{|z_i^0 - z_j^0 |^3}
\label{eq:BHMhamiltonianRadial}
\end{equation}
Introducing the quantities
\begin{equation}
\begin{array}{lcl}
t_{i,j} & = & \displaystyle{\frac{1}{2} \frac{1}{4\pi\epsilon_0}\frac{e^2}{m \nu_{x,1}^2}\frac{\hbar \nu_{x,1}}{|z_i^0 - z_j^0|^3}}\\[5mm]
\nu_{x,1}^{(i)} & = & \nu_{x,1} - \displaystyle{\sum_{j=1, j\neq i}^{N} t_{i,j}}
\end{array}
\label{eq:hoppingAndTrapping}
\end{equation}
the second quantized form becomes \cite{Deng2008}
\begin{equation}
\frac{H_{x0}}{\hbar} = \sum_{i=1}^{N} \nu_{x,1}^{(i)} n_i + \!\!\!\!\!\!\! \sum_{i,j=1, i>j}^{N} \!\!\!\!\! t_{i,j}\left(a_i^{\dagger} + a_i\right)\!\!\left(a_j^{\dagger} + a_j\right)
\end{equation}
The quantities $t_{i,j}$ are site dependent hopping terms resulting from the non vanishing coupling of the radial motion between all ions in the chain. Coupling of ion motions and therefore tunneling of transverse phonons is the stronger the closer the ions are. The next-neighbour tunnelling is therefore the strongest but its slowly decaying behaviour with lattice site difference can already be seen from equation (\ref{eq:BHMhamiltonianRadial}), leading to long range tunnelling in stark contrast to pure next-neighbour tunnelling, as found in neutral atom experiments.
The Coulomb repulsion from adjacent ions has a maximum for $\delta x=0$ and counteracts the radial trapping potential, and the $\nu_{x,1}^{(i)}$ are the reduced radial trapping frequencies. These changes in the radial trapping frequencies and therefore the reductions in the phonon energies are the strongest for small ion separations and thus lead to an axial phonon confinement towards the centre of the trap \cite{Deng2008}. This confinement becomes the stronger, the more inhomogeneous the ion separations are. Therefore this effect is more pronounced in a linear ion trap compared to the isospaced array of microtraps (with almost constant ion separations) and becomes more evident with increasing ion number.

\begin{figure}[t!]
    \centering
        \includegraphics[width=1.0 \columnwidth]{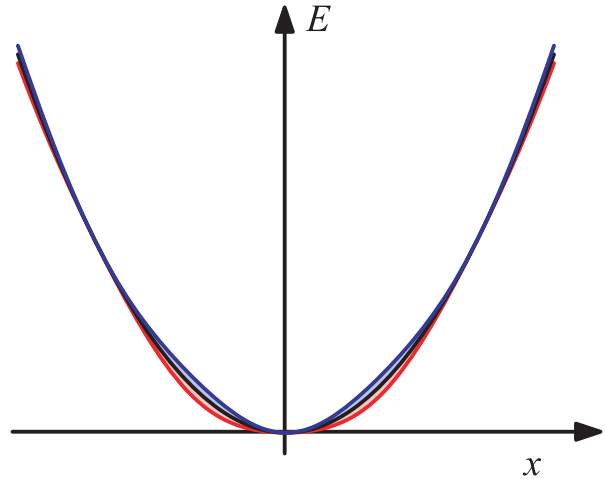}
    \caption{(colour online) The transverse potential as a sum of a parabolic trapping potential plus a far off-resonant standing wave: the black curve shows the unperturbed effective trap potential, while the red is the total effective potential with a standing wave having a maximum at the ion position, resulting in a effective repulsive interaction of phonons. The blue shows the total effective potential with a standing wave having a node at the ion position, resulting in a effective attractive interaction of phonons.}
    \label{fig:BHMpotential}
\end{figure}
\begin{figure}[t!]
    \centering
        \includegraphics[width=1.0\columnwidth]{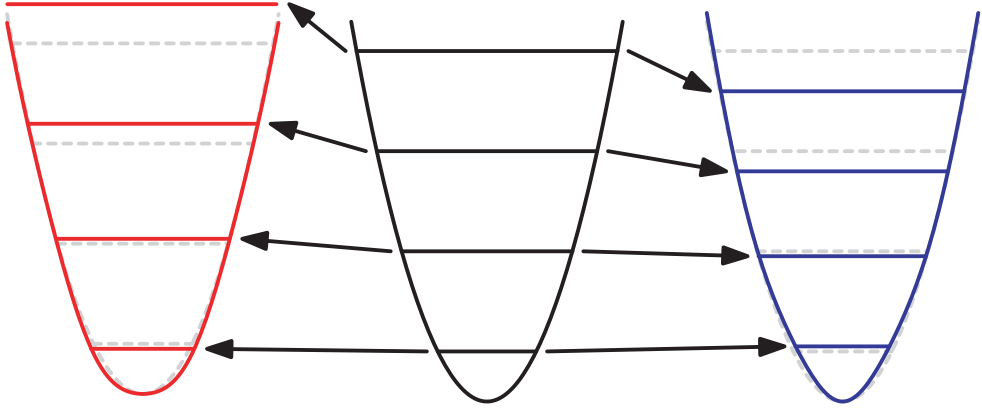}
    \caption{(colour online) Shifts of the harmonic oscillator levels due to the standing wave (not to scale). The centre image shows an undistorted harmonic trap with equally spaced energy levels. When a standing wave along $x$ is added, such that a maximum is located at the trap centre (left hand side), the base level is lowered due to the quadratic part in the standing wave. In addition, the higher levels are raised due to the quartic part of the dipole potential. The levels spacing is no longer equidistant and mimics a repulsive interaction of the phonons. For an additional standing wave with a minimum at the trap centre (right hand side), the lowest level is raised due to the upward curvature of the dipole potential. The level spacing is reduced and also non equal and pretends a attractive interaction of the phonons. The grey dashed lines in the outer graphs again indicate unperturbed potential and energy levels.}
    \label{fig:BHMlevels}
\end{figure}

Due to the large separation of ions and the small transverse oscillation amplitudes, the effect of the Coulomb interaction to the radial potential is just a change in the trapping frequency and the anharmonicities, that should in principle be already present, have been neglected so far. A purely harmonic trapping potential results in equally spaced oscillator levels, meaning that the energy of having two phonons located on different sites or both at the same site gives the same total energy. Interacting phonons means that the energy of having two phonons at the same site is the sum of the the two single phonon energies, altered by an interaction energy, and the oscillator levels are no longer equidistant. A nifty idea to mimic this interaction is to introduce an anharmonicity to the trapping potential and thus to break the regular spacing of the trap levels. This can in principle be done by any non harmonic potential as, for example a sinusoidal potential as known from optical lattices generated by an off-resonant standing wave with wave vector $k_{\rm{SW}}$ \cite{Deng2008}:
\begin{equation}
H_{\rm{SW}} = F \sum_{i=1}^N \cos^2\left(k_{\rm{SW}} \; x_i + \frac{\pi}{2}\delta\right)
\label{eq:standingwave}
\end{equation}
where $F$ is the peak AC Stark energy shift. For small anharmonicities the Hamiltonian $H_{x0}+H_{SW}$ can be expanded in the Lamb-Dicke-parameter $\eta_{x,\rm{SW}}$ with
\begin{equation}
\eta_{x,\rm{SW}} =  \sqrt{\hbar^2 k_{\rm{SW}}^2 / 2 m \hbar \nu_{x,1}}
\label{eq:lambbdickeSW}
\end{equation}
and is sufficiently approximated when truncated after second order terms and the Hamiltonian can be brought into the well known form of the Bose-Hubbard model \cite{Deng2008,Fisher1989}:
\begin{equation}
\begin{array}{rl}
H_x^{B\!H\!M} =  & \displaystyle \sum_{i,j=1, i>j}^{N} \hbar\; t_{i,j} \left(a_i^{\dagger} a_j + H.c. \right)\\[5mm]
& + \displaystyle \sum_{i=1}^{N} \hbar \left(\nu_{x,1}+\nu_{x,1}^{(i)}\right) a_i^{\dagger} a_i \\[5mm]
& + U \sum_{i=1}^{N} a_i^{\dagger 2} a_i^2  \end{array}
\label{eq:Bose-Hubbard}
\end{equation}
In this approximation of small anharmonicities, the total radial phonon number is conserved and coupling to axial modes is neglected. The third term in equation \ref{eq:Bose-Hubbard} has a contribution proportional to the phonon number squared which is responsible for the interaction. Its strength is scaled by \cite{Deng2008}
\begin{equation}
U = 2 (-1)^{\delta} \, F \eta_{x,\rm{SW}}^2
\label{eq:interaction}
\end{equation}
The interaction depends on the strength and the relative position $\delta$ of the standing wave and can even change its sign (see equations \ref{eq:standingwave} and \ref{eq:interaction}): it is maximum positive/repulsive for ions at the maxima of the standing wave $(\delta = 0)$ and maximum negative/attractive for ions at the minima of the standing wave $(\delta = 1)$. This can be seen in figure \ref{fig:BHMpotential} and \ref{fig:BHMlevels}: the small sinusoidal alteration from the standing wave makes the tip of the parabola slightly pointier for a minimum of the standing wave or chamfered for a maximum of the standing wave which results in a shift of the trapping frequency. The interesting consequence, however, are the differential shifts of the oscillator levels which are depicted in figure \ref{fig:BHMlevels}: the fourth order contribution from the extremum of the sinusoidal potential has always opposite sign of the quadratic contribution. Thus it is curved upwards for a maximum of the standing wave (meaning that the quadratic term is curved downwards) and shifts higher levels up, mimicking a repulsive interaction. A minimum of the standing wave has an upwards curved quadratic contribution; the fourth order is curved downwards, making the trap softer when walking away from the centre and reduces the energy of higher trap levels imitating an attractive interaction.

The Bose-Hubbard model is well known in neutral atom physics \cite{Fisher1989}. When the total phonon number $N_{\rm{ph}}$ is an integer multiple of the ion number $N_{\rm{ion}}$ the model predicts a transition from the superfluid state (dominated by tunnelling) to a Mott-insulator state (negligible tunnelling) with a constant phonon number per lattice site. This transition can be traversed by  changing the ratio of tunnelling frequency and repulsive interaction energy (that is, changing trap frequency or standing wave intensity). For incommensurate phonon numbers that are no integer multiple of the ion number, a finite number of stages with constant phonon occupation can be anticipated, with the highest phonon populations at the bottom of the phonon confining potential, that is, at the trap centre. In neutral atom physics, the superfluid-Mott insulator transition has been observed in beautiful experiments \cite{Greiner2002}.

The framework of Luttinger liquid theory \cite{Luttinger1963,Voit1995} predicts algebraic decay of correlations between the number of phonons for the superfluid phase when phonon tunnelling beyond next-neighbours is neglected \cite{Deng2008}
\begin{equation}
C_{i,j}^{nn} = \left\langle n_i n_j \right\rangle - \left\langle n_i\right\rangle \left\langle n_j \right\rangle \propto |i-j|^{-2}
\label{eq:Luttinger-correlationDiagonal}
\end{equation}
and correlations that are nondiagonal in the phonon number decay as
\begin{equation}
C_{i,j}^{aa} =\frac{\left\langle a_i^\dagger a_j \right\rangle}{\sqrt{\left\langle n_i\right\rangle \left\langle n_j \right\rangle}} \propto |i-j|^{-\alpha}
\label{eq:Luttinger-correlationNonDiagonal}
\end{equation}
with $\alpha \propto \sqrt{U/t\, n_0}$ given by the ratio of tunnelling and interaction at the central phonon density $n_0$. In the Mott phase, however, both correlations decay exponentially with distance:
\begin{equation}
C_{i,j}^{aa,nn} \propto e^{-|i-j|/\xi}
\label{eq:Mott-correlation}
\end{equation}
with the correlation length $\xi$.

Phonon densities and fluctuations from numerical calculation using the density matrix renormalization group method (DMRG) \cite{White1993} were presented in \cite{Deng2008} for linear traps and an array of microtraps. The calculations reproduced the superfluid Mott-insulator transition starting from the sides of the chain (due to larger ion separations, the tunnelling rate is lower at the ends). Especially for an array of microtraps the Mott phase features an almost flat mean phonon number per lattice site along the chain for commensurate ion and phonon number and the fluctuations are also flat and strongly reduced. For linear traps the inhomogeneous ion separations modulate phonon numbers and fluctuations which are both more peaked at the chain centre. The correlations in the superfluid phase are found to be in agreement with (\ref{eq:Luttinger-correlationDiagonal},\ref{eq:Luttinger-correlationNonDiagonal})) for short (linear trap) and intermediate (microtrap array) distances and signatures of finite size effects and inhomogeneous ion separations can be seen and are found to be more important in linear traps. In addition, long range correlations in Mott phase are found that can be attributed to the long range hopping terms in equation (\ref{eq:hoppingAndTrapping}). For large repulsive interactions and incommensurate filling the system was found in \cite{Deng2008} to form a Tonks-Girardeau-gas \cite{Girardeau1960,Lieb1963} which has been recently observed in neutral atoms \cite{Paredes2004,Tolra2004}. Due to their repulsion, the phonons cannot get past each other and attain the hard core bosons limit and tunnelling is suppressed. The atoms/phonons behave neither purely bosonic nor fermionic, because they can occupy the same momentum state but not the same position in space. Mathematically, there is an exact one-to-one mapping between impenetrable bosons and non interacting fermions for a one-dimensional system. Numerical studies in \cite{Deng2008} for an array of microtraps and half filling ($N_{\rm{ph}} = 1/2~ N_{\rm{ion}} $) show a constant phonon number distribution together with algebraically decaying correlations. The exponent of the decay deviates from the value $1/2$ which is expected for a Tonks gas with nearest neighbour interaction only. This is explained by mapping the BHM to a $XY$ model with antiferromagnetic dipolar interactions. \cite{Deng2008} also study attractive systems 
and use site dependent interactions to construct systems with highly degenerate ground states described by an $XY$ model.

\subsection{Experiments}
\label{BHMExperiments}

For a string of 50 ions with a minimum axial separation of 5~\textmu m and a radial trapping frequency $\nu_{x,1}$ about 70 times higher than the axial trapping frequency $\nu_{z,1}$ the tunnelling rates $t_{ij}$ were found to be comparable to the axial trapping frequency. For $F=\hbar\nu_{x,1}$ which is realistic with commercially available laser systems, the axial frequency is changed by a few ten percent and the system remains in the phonon number conserving regime and the phonon interaction energy $U$ is approximately twice the tunnelling rate \cite{Deng2008}.

The interesting quantities are the number of atoms at each lattice site, corresponding to the number of transverse phonons at each ion, their fluctuations and their correlations. Phonon occupation can not be measured directly but can be mapped on the internal state and the ions' resonance fluorescence is recorded. The mapping is accomplished by sideband-transitions with carrier Rabi frequency $\Omega$ -- note that a sideband doesn't refer to the axial mode as usual but to the radial mode along the $x$ axis here    . Since the strength of the sideband transition depends on the expectation value of the phonon number, the transition matrix element, and therefore the vibrational state can be extracted from repetitive measurements \cite{Meekhof1996A,Meekhof1996B,Leibfried1996}. If the ion is not in a Fock state, each phonon population contributes to a $\sin^2(\sqrt{n} \Omega t)$ term in the temporal evolution, which is accordingly given by
\begin{equation}
    P_{\uparrow} = \sum_n P(n) \sin^2(\sqrt{n} \Omega t)
    \label{eq:phonon_evolution}
\end{equation}
with $P(n)$ being the probability of having $n$ phonons. Different frequency components in the temporal evolution can be extracted from Fourier analysis to yield the $P_n$ \cite{Meekhof1996A,Meekhof1996B,Leibfried1996} and reveal directly and transition from superfluid to Mott insulator state: For a Mott insulator, all ions are in the same vibrational state and therefore have the same transition matrix element (assuming the Number of phonons being an integer multiple of the ion number). Exciting sideband transitions on all ions simultaneously brings all of them in the same internal state, so for appropriately chosen ($\pi$-pulse) times the whole chain will switch from dark to bright. For a superfluid, however, the phonon expectation value is not a constant along the chain, and there is no universal $\pi$-pulse and the ions will never undergo common oscillations from the dark to the bright state when varying the interaction time. If desired, the whole density matrix can be deduced from quantum state tomography which, of course, becomes tedious or, for all practical purposes, even impossible for long chains.

\subsubsection{Superfluid-Mott insulator transition and creation of a superfluid phonon state by adiabatic evolution}
\label{sec:SuperfluidMottInsulatorTransition}

This experiment is similar to the superfluid-Mott insulator transition by \cite{Greiner2002} obtained by ramping the optical lattice depth. A  clear demonstration, as proposed in \cite{Porras2004A} would start with an ion chain cooled to a state with zero radial phonons, the "radial ground state". The optical lattice is ramped up to a high intensity $U \gg t_{ij}$ such that the ions sit on a node of the standing wave and the phonons experience a repulsive interaction. A defined number of phonons is introduced by sideband transitions, such that the total phonon number is an integer multiple of the ion number (for example by a $\pi$ pulse on the blue sideband). If now the phonon number is mapped back to the internal state by sideband transitions, a collective switching from the internal bright to the non scattering state should be observed, that is, the time for a $\pi$ pulse is identical throughout the whole chain). Alternatively, the interaction $U$ is adiabatically reduced down to a final strength $U_f$ (the standing wave intensity is lowered), such that the system remains in the ground state. At a critical value $U_f \approx t_{ij}$ the system undergoes a transition to a phonon superfluid. When measuring phonon occupation numbers by mapping them to internal states by sideband transitions, the phonon numbers can be expected to be inhomogeneous, due to the phonon trapping potential and non negligible tunnelling.

\subsubsection{Bose-Einstein-condensation by evaporative laser cooling}
\label{sec:BECByEvaporativeLaserCooling}

Evaporative laser cooling for phonons in ion traps sounds like an oxymoron because in neutral atom physics evaporative cooling terms a process free of laser scattering, where only the highest energy atoms can leave the trap \cite{Davis1995A} 
and take more than the average thermal energy with them which, provided sufficient fast thermalization) results in a cooler cloud. The analogy is, that in the ion trap with an axial phonon confining potential high phonon occupations (labelled high energy phonons in \cite{Porras2004A}) can more easily tunnel to the ends of the chain, analogous to hot atoms, and when laser cooling is applied to these ends exclusively, the phonons are removed or evaporated.

A demonstration would start with a Doppler cooled ion chain with a given number of phonons per site. Applying laser cooling at the ends of the ion chain removes or evaporates high energy phonons (or phonon occupation numbers) from the top of the phonon confining potential. A small interaction $U \ll t_{ij}$ allows for a phonon-phonon interaction and therefore for thermalization, equivalent to the need of a non vanishing scattering length for thermalization of neutral atom clouds and finally Bose-Einstein condensation of phonons can be observed. Since this is a 1D system, BEC cannot take place in the thermodynamic limit $N\rightarrow\infty$ at finite temperature, but for a given in number there will be a non-zero transition temperature. In contrast to the Mott insulator (phonon Fock state), the signature of this phase transition is more difficult to extract. One can deduce phonon occupation distribution from series of fluorescence images after sideband transitions of varying time. Evaporative laser cooling for phonons should reveal an abrupt transition from a thermal phonon occupation distribution to the BEC distribution, which corresponds to a distinction of a Gaussian and a Thomas-Fermi momentum distribution for neutral atoms.

It could be interesting to modify the axial trapping potential in microtraps such that the equilibrium positions are really equidistant and the phonon trapping potential, probably corresponding to unconfined phonons and a demonstration Bose-Einstein-condensation of free particles, similar to experiments by \cite{Meyrath2005}.

\subsection{Frustrated XY models}
\label{sec:frustratedXYmodels}

For large anharmonicities,  strong repulsion, and low filling factor $N_{\rm{ph}} \ll N_{\rm{ion}}$ the occupation by two phonons becomes unlikely and the system attains the hardcore Boson limit. The phonons can be mapped to $S=\frac{1}{2}$ spins via the Holstein-Primakoff transformation as ($a_\alpha^\dagger \rightarrow S_\alpha^+$, $a_\alpha \rightarrow S_\alpha^-$ and $n_\alpha \rightarrow S_\alpha^z + \frac{1}{2}$). The resulting Hamiltonian is an XY model with dipolar interactions and a site dependent potential \cite{Schmied2008}:
\begin{equation}
H_S = 2 \sum_{\left\langle\alpha,\beta\right\rangle}    t_{\alpha,\beta} \left(S_\alpha^x S_\beta^x + S_\alpha^y S_\beta^y \right) + \sum_\alpha V_\alpha S_\alpha^z
\end{equation}
Hamiltonians like theses are interesting because they allow to investigate frustrated XY models in one and two dimensions.  Frustration is a phenomenon relevant in condensed matter physics and describes the situation where, due to the lattice geometry or competing interactions, the interaction energies cannot be simultaneously minimized which can result to highly non-degenerate ground states with non-zero entropy, even at zero temperature.
Such frustrated XY models can now  be investigated in 1D and 2D system, that is, in linear chains and planar zigzag structures in linear ion traps and planar structures in surface traps.

The 1D case leads to the $S=\frac{1}{2}$ XY antiferromagnet with dipolar interactions, which becomes exactly solvable if only next neighbour interactions are considered. By weakening the transverse confinement along one direction the linear chain is no longer the energetically favoured pattern and the ion arrange themselves in a planar zigzag ladder which can be viewed as two linear chains relatively displaced to each other. By variation of the radial trapping potential, the properties of this two-dimensional structure, namely the zigzag amplitude and thus the ratio between the dominant inter ($t_1$) and intra-chain ($t_2$) coupling can be tuned. The zigzag amplitude $\xi$, measured in units of the inter ion spacing, is limited by the fact that for a too weak transverse confinement the ion chain spontaneously flips into helical order (around $\xi \approx 0.965$). This in turn limits the accessible range of coupling ratios when phonons perpendicular to the zigzag plane are considered. But when exploiting phonons in the ladder plane any $|t_2 /t_1|>1/8$ ratio can be accessed. For low zigzag amplitudes and intra-chain couplings, the ladder is found to exhibit long range antiferromagnetic N\'{e}el order and undergoes a transition at $\xi\approx0.461$ to spiral order which shows both spontaneous magnetic order and chiral order.

The chirality can be defined as
\begin{equation}
    \kappa_i = 4\left( S_i^x S_{i+1}^y - S_i^y S_{i+1}^x\right)
\label{eq:chirality}
\end{equation}
and the chiral order parameter is the averaged chiral correlation over all ion-pairs in the ladder:
\begin{equation}
\begin{array}{rcl}
    O_\kappa^\Delta & = & \displaystyle \frac{1}{L - 1 - |\Delta|}\sum_i \left\langle \kappa_i \kappa_{i+\Delta}\right\rangle \\
    O_\kappa & = & \displaystyle \frac{1}{2L-3} \sum_{\Delta = -(L-2)}^{L-2} \!\!\!\!\!\!\!  O_\kappa^\Delta
\end{array}
\label{eq:chiralOrder}
\end{equation}
When envisioning an experiment measuring this chiral order, we first have to remember that the spins in \ref{eq:chirality} are not real spins but phonons mapped back via the Holstein-Primakoff transformation and that the probability of having double phonon occupation is negligible. The expectation values for the phonon occupation are thus given by the probability of having single phonon occupation. To measure chirality and the chiral order parameter, we have to measure phonons in both transverse modes for each ion and their correlations by sideband transitions. When the measurement is repeated for different radial confinements and thus coupling ratios the transition from N\'{e}el to sprial order can be observed. Another interesting feature to look at is a yet not understood ``reorientation transition'' transition around $\xi \approx 0.8$ found in numerical calculations of up to 20 ions \cite{Schmied2008} which gets more pronounced for larger ion numbers. Furthermore, the convergence of the numerical algorithm used in \cite{Schmied2008} shows poor convergence for equal inter and intra chain couplings and thus a measurement of the phonon correlation function in this regime could be interesting.

Even more variations are possible in extended planar lattices, as, for example triangular lattices as considered in \cite{Schmied2008}. Rotating planar triangular Wigner crystals have been observed in Penning traps \cite{Mitchell1998} and again they can be viewed as displaced linear chains such that it is possible to define a ratio of inter and intra chain coupling. A variety of phases has been found when the orientation of the a co-rotating standing wave, and therefore vibrational motion is changed relative to the lattice normal \cite{Schmied2008}. Another approach to implement such lattices is by means of surface traps \cite{Chiaverini2008,Seidelin2006}. The equilibrium ion positions and therefore the ratio of inter and intra chain couplings is given by the effective potential generated by the trap. A variation of the ratio of relevant couplings could be investigated either in a series of surface traps, each designed for a particular coupling ratio, or by electrode geometries which allow for a anisotropic deformation of the triangular lattice. It should be pointed out that such planar traps would in principle allow to implement any desired lattice geometry.

\section{Axial standing wave: The Frenkel-Kontorova-Ion-Model}
\label{sec:FKIM}

The Frenkel-Kontorova-Model (FKM) is a model describing a linear chain of particles with harmonic next neighbour interactions which is exposed to a sinusoidal potential \cite{Frenkel1938}. The system of interest might remind the reader of the Hubbard- or Bose-Hubbard-model as discussed in section \ref{sec:BHMBackground}, however, the Frenkel-Kontorova-Model is simpler, being classical and one-dimensional, and, on the other hand, focusses more on classical nonlinear dynamics and chaos, and questions of integrability and the persistence of quasi-periodic motion \cite{Braun1998}. Despite this simplicity the FKM is able to describe a variety of physical phenomena as crystal dislocations, commensurate-incommensurate phase transitions, epitaxial monolayers on a crystal surface, magnetic chains, and fluxon dynamics in Josephson junctions (see \cite{Braun2004}, and references therein). The FKM is one of the first examples in solid state physics where a one dimensional chain can be used to model an extended two dimensional defect in a bulk \cite{Braun1998}.

\subsection{Formalism and numerical calculations}
\label{sec:FormalismAndNumericalCalculations}

\begin{figure}[t!]
    \centering
        \includegraphics[width=0.9\columnwidth]{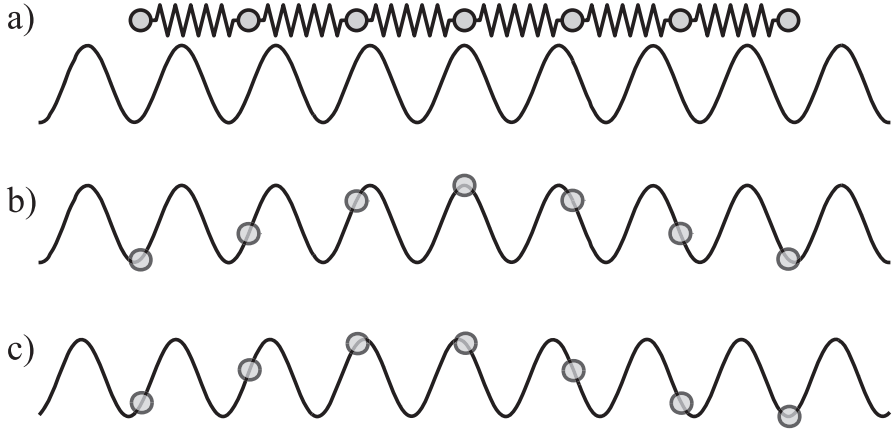}
    \caption{a) basic Frenkel Kontorova model: a one dimensional chain of atoms with next neighbour interactions is exposed to a sinusoidal potential; figures b) and c) show possible spatial configurations: the upper configuration b) is unstable -- the total energy, which is a function of the coordinates of all particles has a saddle point, whereas configuration c), obtained only by a minuscule sideways shift is stable, corresponding to a minimum of the total energy (after \protect\cite{Braun1998})}
    \label{fig:FKMprinciple}
\end{figure}

In the basic model the interactions are harmonic and restricted to next-neighbours and the external potential is sinusoidal (see figure \ref{fig:FKMprinciple}). Without this external potential the dynamics of the chain can be understood in terms of normal modes and is quasi-periodic. When the potential is switched on but the perturbation is small, that is, below the critical perturbation strength, the system is in the so called \textit{sliding phase} and the chain can oscillate in the lattice and the ion positions follow a Kolmogorov-Arnol'd-Moser curve \cite{Braun2004}.  For higher perturbations, the system undergoes an \textit{Aubry analyticity breaking transition} \cite{Aubry1983} and the ion chain starts to be pinned by the lattice and the positions form a devils staircase corresponding to a fractal Cantor set \cite{Garcia-Mata2007}. The ground state is unique for all perturbations, but in the pinned phase, a large number of states exists very close to the ground state like in a fractal spin glass \cite{Zhirov2002}.

Various extensions and modifications of the basic Frenkel-Kontorova model have been investigated, for example arbitrary on site potentials and anharmonic interactions between the particles \cite{Braun1998}. In the following we will focus on the Frenkel-Kontorova-Ion-Model (FKIM) which is a classical picture for a laser-cooled chain of ions trapped in a linear trap exposed to a standing wave \cite{Garcia-Mata2007}. For a sample of ions in a linear trap with a much stronger radial than axial confinement, the ions arrange themselves in a linear chain. When the radial motion is frozen due to laser cooling, the chain can be described in good approximation with a one dimensional Hamiltonian. Then the interaction of the particles is given by the long range Coulomb force and the external potential is a sum of the effective sinusoidal potential from the standing wave and the effective  harmonic axial trapping potential. The Hamiltonian is given by
\begin{equation}
 \begin{array}{rl}
H = \displaystyle\sum_{i=1}^{N} & \left( \displaystyle{\frac{p_{z,i}^2}{2m} + \frac{m \,\nu_{z,1}^2}{2}z_i^2 - K \cos(2\pi  z_i/d)}\right)\\
& + \displaystyle{\sum_{i>j} \frac{e^2}{4\pi\epsilon_0} \frac{1}{\left| z_i - z_j\right|}}.
\end{array}
\label{eq:FKIM_Hamiltonian2}
\end{equation}
The Hamiltonian above corresponds therefore to the general ion trap Hamiltonian given in equation (\ref{eq:ion_chain_Hamiltonian}) where only the part in $z$-direction is considered and an sinusoidal potential of strength $K$ and periodicity $d$ is added. A basic FKM Hamiltonian can be obtained by neglecting restricting the interaction sum to next neighbours and expanding the sum of interaction and trapping potential in the deviations from the equilibrium positions up to second order. In \cite{Garcia-Mata2007} the authors investigate the dimensionless form of the FKIM-Hamiltonian above which reads
\begin{equation}
H = \sum_{i=1}^{N} \left( \frac{P_i^2}{2} + \frac{\tilde{\nu}^2}{2}\tilde{z}_i^2 - \tilde{K} \cos \tilde{z}_i\right)
+ \sum_{i>j} \frac{1}{\left| \tilde{z}_i - \tilde{z}_j\right|}.
    \label{eq:FKIM_Hamiltonian}
\end{equation}

The ion positions $\tilde{z}$ are given in units of the reduced lattice constant $\tilde{d}=d/2\pi$ and the energy $\tilde{E}$ as well as the strength of the sinusoidal potential $\tilde{K}$ are measured in units of $e^2/4\pi \epsilon_0 \tilde{d}$. The modified angular frequency is obtained as  $\tilde{\nu}^2 = 4\pi \epsilon_0 m \nu_{z,1}^2 \tilde{d}^3/e^2$. In the quantum case the momenta are given by $P_i = -i \hbar_{\rm{eff}} \partial_{x_i} $ with an effective Planck constant $\hbar_{\rm{eff}}$ given as
\begin{equation}
\hbar_{\rm{eff}} = \hbar / \left(e\sqrt{md}\right)
\label{eq:effHBar}
\end{equation}
which depends on the lattice constant which allows to scale and investigate quantum effects as we will see later.
\begin{figure}[t!]
    \centering
        \includegraphics[width=1.0\columnwidth]{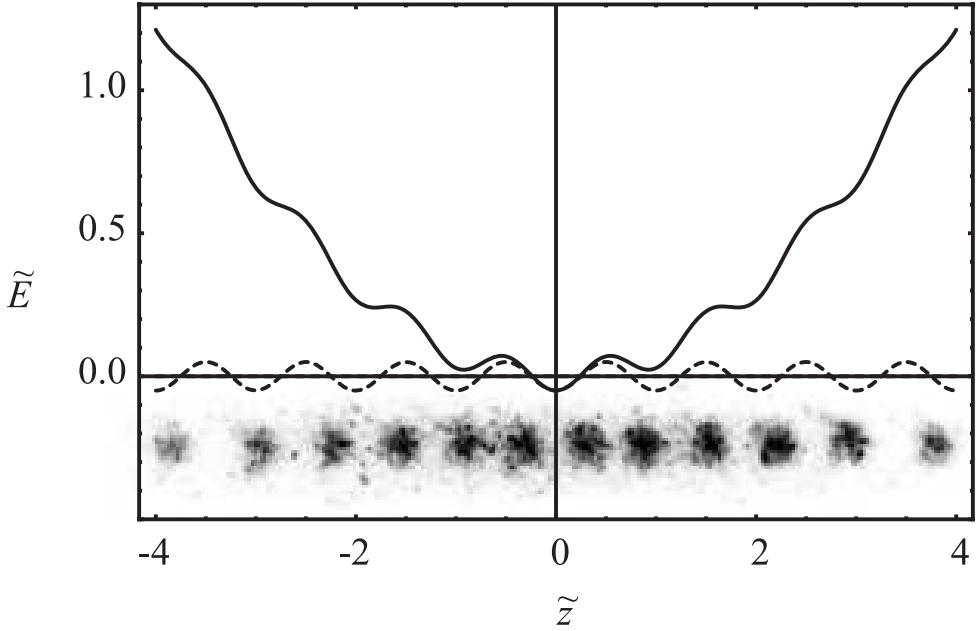}
    \caption{Frenkel-Kontorova ion model: The potential curvature, the sine ripple amplitude and lattice constant are plotted for values corresponding to the \emph{Aubry analyticity breaking transition} at golden mean density for a chain of 12 ions with a centre separation of 5 \textmu m. A CCD image of a string of 12 $^{172}$Yb$^+$ions is scaled to match the conditions for the potential plot and indicates the unperturbed ion positions.}
    \label{fig:FKIMpotential}
\end{figure}

First the authors compare numerical solutions to the classical ground state problem in the FKIM to solutions of the FKM with only next neighbour interactions. The numerical studies are performed for up to a few hundreds of ions at the golden mean density (at the center of the ion chain), where the average occupation of a lattice site is $(\sqrt{5}+1)/2$ and results in the golden Kolmogorov-Arnol'd-Moser (KAM) curve for the ion positions commonly used for studies of the Aubry transition \cite{Aubry1983,Braun2004}. The FKM leads to dynamical recursive maps for the calculation of equilibrium ion positions which are found to describe the centre 1/3 part of the ion chain appropriately. By comparison of the solutions for FKM and FKIM equilibrium positions, the influence of long range interactions and the trapping potential can be mapped out and is found to be negligible in the centre region. Here, the ion density is highest and almost constant and allows for screening of the trap potential by nearby ions, similar to the Debye radius in plasma physics. In the outer regions visible deviations in the equilibrium positions are found, as the ion density is lower at the ends of the chain.

As for the FKM, for small depths of the sinusoidal potential ($\tilde{K}<0.05$), the chain is in a sliding phase. The hull function, which compares the unperturbed equilibrium positions to the perturbed ones modulo $2\pi$ is continuous and the phonon mode spectrum has a sound like form. At a critical strength of the standing wave ($K_c\approx 0.05$), the system undergoes an \textit{Aubry analyticity breaking transition} \cite{Aubry1983}.
Above the critical lattice strength, the ion chain is in a pinned phase, their positions are described by a devils staircase, which is a fractal cantor set \cite{Peitgen2004}, and the phonon spectrum shows a gap. Another similarity to the FKM is that in the pinned phase, the FKIM has properties of a spin glass, and features an enormous number of stable equilibrium configurations energetically close to the ground state \cite{Zhirov2002}, despite the ground state being unique.

So far, all considerations have been purely classical. Quantum effects were in investigated in \cite{Borgonovi1989} and the quasi-degenerate configurations close to the ground state are found to become important and tunnelling between them leads to non-trivial \textit{instanton excitations} \cite{Garcia-Mata2007}.  The quantum case is studied in \cite{Garcia-Mata2007} using the Quantum-Monte-Carlo (QMC) approach \cite{Grotendorst2002,Nightingale1998}. To see this structural changes, the authors look at the power spectrum of the position distribution (in the classical case, the $z_n$ are the ion positions):
\begin{equation}
F(k) = \frac{\left\langle \left| \sum_n  \exp\left(ikz_n (\tau)\right)\right|^2\right\rangle}{\delta N   }
\label{eq:formfactorcorrect}
\end{equation}

From these numerical solutions a phase transitions from a pinned instanton glass to a sliding phonon regime is observed, depending on the value of the \emph{effective} Planck constant $\hbar_{\rm{eff}}$ \cite{Garcia-Mata2007} which can be varied by changing the lattice constant and thus the influence of tunnelling, and the transition taking place at $\hbar_{\rm{eff}} \approx 1$. For low values of $\hbar_{\rm{eff}}$ the formfactor shows discrete resonances which are equidistant in the sliding phase, and and integer $k$ showing that the ions are pinned by the optical lattice. The pinned phase also shows a quantum transition when $\hbar_{\rm{eff}}$ is increased: for low values $\hbar_{\rm{eff}}$, the pinned phase is not destroyed and discrete resonances are still found. For larger values of $\hbar$ the formfactor bacomes continuous and the transition occurs at a value $\hbar_{\rm{eff}} \approx 1$.

\begin{figure}[t!]
    \centering
        \includegraphics[width=1.0\columnwidth]{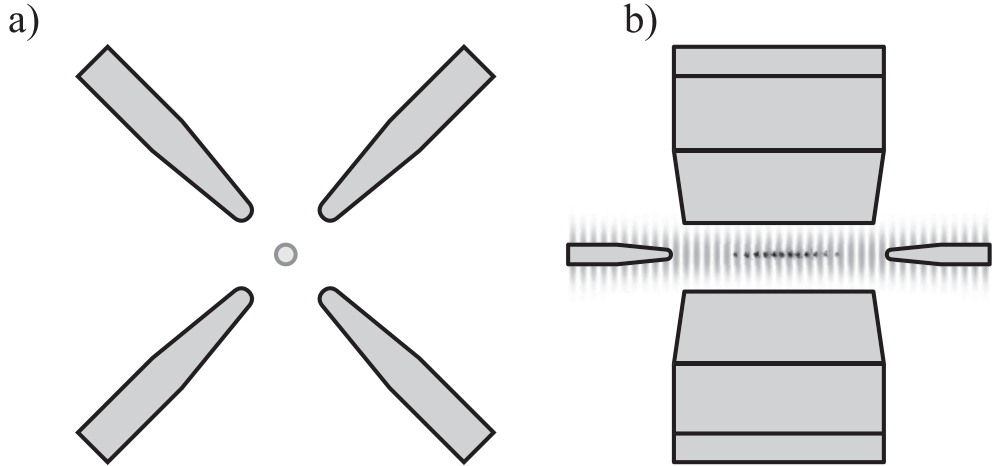}
    \caption{Schematic (not to scale) of the experimental setup suitable for simulating Frenkel Kontorova physics and observing an Aubry transition. The standing wave is applied in the axial direction, either by optical access through ring shaped endcap electrodes, or by crossing two beams at an angle, such that the counter-propagating component is aligned parallel to the weak trap axis. Part a) shows the view along the ion chain, part b) shows the setup as seen by an imaging detector used for measuring the ion fluorescence and for the determination of ion positions and formfactor}
    \label{fig:FKIMsetup}
\end{figure}

\subsection{FKIM Experiments}
\label{sec:FKIMExperiments}

Experiments for investigating Frenkel-Kontorova-physics require a sinusoidal potential along a one-dimensional ion chain. This can be implemented by a far off-resonant standing wave along the weak trap axis as indicated in figure \ref{fig:FKIMsetup}. Being far off resonant means that changes of the internal state due to this lasers is unlikely within the duration of the xperiment. Following \cite{Garcia-Mata2007} the ion density at the centre should be the golden mean density, that is, the lattice periodicity should be by a factor $(\sqrt{5}+1)/2$ larger than the ion separation at the middle of the chain. This can easily be obtained even at a fixed wavelength changing the angle between the intersecting beams that form the dipole potential or, alternatively, by changing the trap frequency. To obtain the golden mean density for ions separated 5~\textmu m, the lattice periodicity is approximately 3.1~\textmu m corresponding for example to two beams at $\lambda = 1064$~nm each enclosing an angle of $\alpha \approx 80^{\circ}$ with the weak trap axis.

The position of the lattice has to be fixed relative to the ions which means that the relative phase of the two counter-propagating running waves has to be constant. This requires stability and potentially active feedback on mirrors. To maintain a constant depth of the sinusoidal potential along the chain, the Rayleigh range or would ideally be large compared to the chain length, meaning that the waist would be large. The required strength of the standing wave $K_c$ to observe an Aubry transitions corresponding to a depth of 0.6~K ions separated by 5~\textmu m which is much deeper than in neutral atom experiments. On the other hands the trap frequencies are much higher in ion trap experiments and thus the time scales for a measurement are shorter, allowing for lasers, that are closer to the atomic resonance and in turn requiring less power. The challenging power requirements can be lessened, when the light cycles in a resonant cavity, however, with a more demanding stability.

The effective Planck constant as given in equation (\ref{eq:effHBar}) contains the lattice constant and can in principle be changed by adjusting either the wavelength of the light that forms the sinusoidal dipole potential or the angle between the intersecting beams. To maintain the golden mean density, the trap frequency would be varied simultaneously. For parameters given above, however, the effective Planck constant is approximately $3\cdot 10{-5}$ (calculated for Ca ions \cite{Garcia-Mata2007}) and due to the weak dependence on the lattice constant (see equation \ref{eq:effHBar}) ion trap experiments are likely to be restricted to the semi-classical regime with $\hbar_{\rm{eff}} \ll 1$, hardly accessible for QMC calculations \cite{Garcia-Mata2007}.

The experiment would start with an ion chain cooled to the ground state. Ramping the standing wave along the weak trap axis adiabatically up to the desired final strength ensures that the system remains in its ground state and the equilibrium positions and thus the formfactor can be measured.

Beside of the general interest in Frenkel-Kontorova physics, an axial standing wave as discussed above might also prove to be useful in the context of quantum information with ion traps since the strong gap in the phonon spectrum found in the pinned phase might allow for the protection of quantum gates against decoherence \cite{Garcia-Mata2007}.

\section{Quantum Fields and relativistic effects}
\subsection{Introduction}

The way to prove a physical theory is to compare its predictions
with the observations. This has been the case with the cosmological
models where effects like the red shift or the homogeneous
background radiation constitute observations in accordance with a
description of an expanding universe and the particle creation. But
to perform a truly cosmological experiment is out of the reach of
our current technology, in particular one that reproduces the early
stages of the universe characterized with an extremely high energy
density and a rapid expansion.

In other cases like with relativistic particles, it is possible to
make experiments were electrons for example are driven to almost the
speed of light. But some interesting effects are still out of reach
of our measurement capabilities. For example the {\it
Zitterbewegung}, predicted from the Dirac equation. It
states that a freely moving spin-1/2 particle, in absence of
external potentials, is subject to a helicoidal motion around the
main direction of propagation. Another counterintuitive effect, the
Klein paradox \cite{Klein1929} states that an electron could
transmit unimpeded through a potential barrier. But this effect has
not been observed yet with elementary particles.




\subsection{Simulating the Unruh effect}

\subsubsection{Methods}

Following the work of Unruh \cite{Unruh1976} it was shown that in
the vacuum of a Minkowski (flat) universe an accelerated observer
would measure a thermal spectrum of particles whose temperature is
proportional to his acceleration. Later this effect was generalized
by Gibbons and Hawking \cite{Gibbons1977} for a curved de Sitter
space. The de Sitter space is a massless curved spacetime that is
isotropic and homogeneous in space, and which looks like the
Minkowski space locally \cite{Bergstrom2004}. The de Sitter space is
also a cosmological model that describes an exponential expansion of
the universe due to a non vanishing and positive cosmological
constant $\Lambda$.

The scalar field describing a massless inflating universe with a
single global mode can be imitated by the quantized motion of a
single ion around its equilibrium position. The phonon excitations
of the ion would simulate the field quanta of the universe. The ion
itself can be used as a detector by coupling its vibrational state
to an easily readable internal state.

In the case of an expanding de Sitter universe a comoving quanta
field detector would experience a Doppler frequency shift of the
form $\nu(t)^\Lambda=\nu^\Lambda_0 \exp{(-\sqrt{3\Lambda} t)}$
\cite{Birrell1984}. The spectrum measured with this detector is
equivalent to the one measured by a thermally excited inertial
detector with a temperature $T=\hbar\sqrt{3\Lambda}/k_B$
\cite{Gibbons1977}.

In the case of a single ion trapped in a time dependent potential
that varies in the form $\nu(t)=\nu_0\exp{(-\kappa t)}$, a similar
effect can be measured. A pseudo-temperature of
$k_BT=\hbar\kappa/2\pi$ is associated with the average occupation
number of the vibrational energy levels.

The coupling between internal and vibrational states of the ion can
be created by a laser field described in the dipole and rotating
wave approximation by the interaction-picture Hamiltonian
\cite{Alsing2005,James1998},
\begin{equation}
H_I=\hbar\Omega k z(t)\left[\sigma^-e^{-i\Delta
t}-\sigma^+e^{i\Delta t}\right].
\end{equation}
Here,  $\sigma^+$ and $\sigma^-$ are the raising and lowering
spin-1/2 operators, $\Delta$  is the detuning of the laser relative
to the atomic transition and $\Omega$ is the Rabi frequency,
respectively. The
laser wave vector $\vec{k}$ is pointing along the direction in which
the ion's vibrational motion is excited (here the $z-$direction). We
assume an anisotropic trap (e.g., a linear trap) where, by choosing
the detuning $\Delta$ and the wave vector $\vec{k}$, the
experimenter can control which vibrational motion to excite.

The evolution of the position operator $z(t)$ is deduced by solving
the equations of motion of the ion in its harmonic trap neglecting
its coupling with the light field. In the Heisenberg representation
and with
 \be
H=\frac{p^2}{2m}+\frac{m}{2}\nu(t)^2z^2 \ ,
 \ee
we have \cite{Menicucci2007}
\begin{equation}
\begin{array}{c}
\displaystyle{\dot{z}=-\frac{i}{\hbar}[H, z]=\frac{p_z}{m}}\\[2 mm]
\displaystyle{\dot{p_z}=-\frac{i}{\hbar}[H, p_z]=\-m\nu(t)^2z}
\end{array}
\end{equation}

After the opening of the harmonic potential $\nu(t)$ during a time
$T$ in which the interaction is on, a vibrational phonon  can be
mapped into the excited $\ket{1}$ internal state. The probability of
exciting the internal state of the ion calculated with first order
perturbation theory and in the slow $\kappa\ll\nu_0$ and long-time
$\nu_0e^{-\kappa T}\ll\kappa$ frequency variation for the trap
potential is \cite{Menicucci2007},
\begin{equation}\label{eq:UnruhProbability}
\begin{array}{ll}
P^{(1)}(\Delta)&=\frac{1}{\hbar^2}\int^T_0dt_1\int^T_0dt_2\bra{0}H_I(t_1)\ket{1}\bra{1}H_I(t_2)\ket{0}\\[2 mm]
&=(\Omega\eta_0)^2\frac{2\pi\nu_0}{\kappa
\Delta^2}\frac{1}{\left(e^{2\pi\Delta/\kappa}-1\right)^2}
\end{array}
\end{equation}

When the trap is opened to simulate the expansion of the universe,
the ion is used as a detector to measure the thermal spectrum from
the phonon distribution. A particular case is when the laser is in
resonance with the red sideband ($\Delta\approx\nu$). The ion acts
like a phonon detector that absorbs the measured particle (together
with a photon) in order to produce an internal excitation of the
ion. When the laser is in resonance with  blue sideband
($\Delta\approx-\nu$), the ion acts like an unconventional detector
that emits a phonon to get excited. The ratio of the excitation
probabilities gives the experimental signature of the ion thermal
distribution \cite{Wineland1979,Monroe1995B},
\begin{equation}
\frac{P_{\rm red}}{P_{\rm blue}}=\frac{\bar{n}}{\bar{n}+1}=e^{-\hbar
\nu_0/k_B T}
\end{equation}
where $\bar{n}$ is the average phonon occupation value. From the
excitation probabilities of equation \ref{eq:UnruhProbability} the
ratio between the red and the blue sidebands is given by
\cite{Menicucci2007},
\begin{equation}
\frac{P^{(1)}(\Delta)}{P^{(1)}(-\Delta)}=e^{-2\pi\Delta / \kappa}.
\end{equation}
Therefore the temperature associated with the exponential opening of
the trap potential is given by $T=\hbar \kappa/2\pi k_B$. This
temperature is equivalent to the Gibbons-Hawking temperature
$T=\hbar\sqrt{3\Lambda}/k_B$ \cite{Gibbons1977}.

The rise of the temperature of the trapped ion is due to the trap
opening. At the beginning of the trap opening the evolution is
adiabatic since the trap frequency obeys the condition
$|\dot{\nu}|/\nu\ll\nu$. Indeed, the expansion constant was selected
small compared to the initial trap frequency $\kappa\ll\nu_0$. At a
later time $T$ chosen following the condition $\nu_0e^{-\kappa
T}\ll\kappa$ the evolution turns to be non adiabatic. Indeed, the
trap frequency is going to be so weak that even a small rate of
change would be non-adiabatic. The ion that initially was in the
ground state is excited to higher vibrational levels because of the
non-adiabatic reduction of the confining potential.

\subsubsection{Experiment}
In this case we measure the temperature associated with the ion
phonon distribution. This temperature is related to the
Gibbons-Hawking temperature.

To perform this experiment a single trapped ion is cooled to the
vibrational ground state level. Next the ion is pumped into its
electronic ground level that is related to the vacuum state of the
universe. Then, the trap potential, initially at the frequency
$\nu_0$, is exponentially reduced $\nu(t)=\nu_0\exp{(-\kappa t)}$.
During the opening of the trap, Raman laser beams between hyperfine
levels or a narrow frequency laser tuned to a quadrupole transition
are set in resonance with the red or the blue sideband. It should be
noticed that the coupling field must be in resonance with the
sideband during the trap opening. After some time, the ion internal
state is measured by fluorescence using the shelving technique with
the help of an additional laser beam in resonance with the excited
internal state.

The trap potential is then raised and the sequence can be repeated
for the opposite sideband. The height of the red and blue sidebands
are compared and the temperature dependence on the trap expansion
coefficient $\kappa$ is measured.

A typical value of the initial trap frequency $\nu_0\simeq1$MHz can
be used to perform this experiment. Therefore the expansion
coefficient can be selected like $\kappa\simeq1$kHz and the
expansion time $T$ of several ms. These are typical parameters in
current ion trap experiments.

Difficulties that may arise when performing such an experiment are
associated with unwanted heating of the ion caused, first, by a
possible unintentional displacement of the trap minimum and by
anharmonic terms in the trap potential, and second, by fluctuating
patch potentials. The former could be solved by compensating the
anharmonicities and the displacement of the trap potential by tuning
appropriate compensating fields for every value of the time-varying
trap frequency. The latter could be overcame by performing the
experiment during a short enough time so that the heating can be
neglected.

\subsection{Simulating the cosmological particle creation}
\subsubsection{Methods}

The standard model describes the particle creation from the vacuum
fluctuations during the early stages of the universe. Since the
expansion of the universe is non adiabatic, an initial ground state
will be excited evolving into a squeezed state of entangled pairs of
particles. The created particle pairs are simulated by entangled
phonon pairs created by the excitation of the ground ionic motional
level as the trap potential is non-adiabatically varied.

One can use a real massless scalar field $\phi$ as a an
approximation to describe an expanding or contracting universe.
Using the Friedman-Lema\^{\i}tre-Robertson-Walker metric
\cite{Birrell1984}, $ds^2=a^6(t)dt^2-a^2(t)d{\bf r}^2$ one finds
that the wave equation describing the universe reads
\cite{Schutzhold2007},
\begin{equation}\label{PartCreField}
\left(\frac{\partial^2}{\partial t^2}+[a^4(t){\bf k}^2+\zeta
a^6(t)\Re(t)]\right)\phi_{\bf k}=0
\end{equation}
where the Ricci (curvature) scalar $\Re$ is coupled to the field via
the dimensionless parameter $\zeta$. Each mode {\bf k} represents a
harmonic oscillator with a time dependent potential $a^4(t){\bf
k}^2+\zeta a^6(t)\Re(t)$.

In the case of a chain of ions trapped in a transversally strongly
confining linear trap with a time dependent axial trap frequency
$\nu_z(t)$, the position $z_i$ of the {\it i}th ion follows the
equation of motion
\begin{equation}\label{eq:ClasIonCristal}
\ddot{z}_i+\nu_z^2(t) z_i=\gamma\sum_{j\neq
i}\frac{\rm{sign}(i-j)}{(z_i-z_j)^2}
\end{equation}
This equation can be solved classically by the use of the scaling
ansatz,
\begin{equation}
z_i(t)=b(t)z_i^0
\end{equation}
where $z_i^0$ are the initial equilibrium positions. The scaling
factor $b(t)$ evolves following
\begin{equation}
\left(\frac{\partial^2}{\partial
t^2}+\nu_z^2(t)\right)b(t)=\frac{\nu_z^2(t=0)}{b^2(t)}
\end{equation}
The quantum fluctuations are introduced by splitting the position
operator $\hat{z}_i$ into its classical evolution $b(t)z^0_i$ and a
quantum fluctuation term $\delta \hat{z}_i(t)$,
\begin{equation}\label{PosIonQuantFluc}
\hat{z}_i(t)=b(t)z^0_i+\delta \hat{z}_i(t).
\end{equation}
Replacing equation \ref{PosIonQuantFluc} in equation
\ref{eq:ClasIonCristal} and linearising later over $\delta
\hat{z}_i(t)$ (since the displacement of the ions can be assumed
small) one obtains after normal mode expansion
\cite{Schutzhold2007},
\begin{equation}\label{PartCreIon}
\left(\frac{\partial^2}{\partial
t^2}+[\nu^2_{z}(t)+\frac{\nu_\kappa^2}{b^3(t)}]\right)\delta
\hat{z}_\kappa=0
\end{equation}
for the phonon modes $\kappa$. One can recognize the similarities
between equation \ref{PartCreField} and equation \ref{PartCreIon}.
The scalar field $\phi_{\bf k}$ behaves in the same way as the
quantum fluctuation of the ions (phonon) $\delta \hat{z}_\kappa$ and
each mode ${\bf k}$ of the field has a direct analogy with the
phonon frequency of the ions $\nu_\kappa$.

It has been shown that a non-adiabatic change in the trap potential
can create squeezed states \cite{Heinzen1990}.

\subsubsection{Experiment}
\begin{figure}[tbh]
    \centering
        \includegraphics[width=0.9\columnwidth]{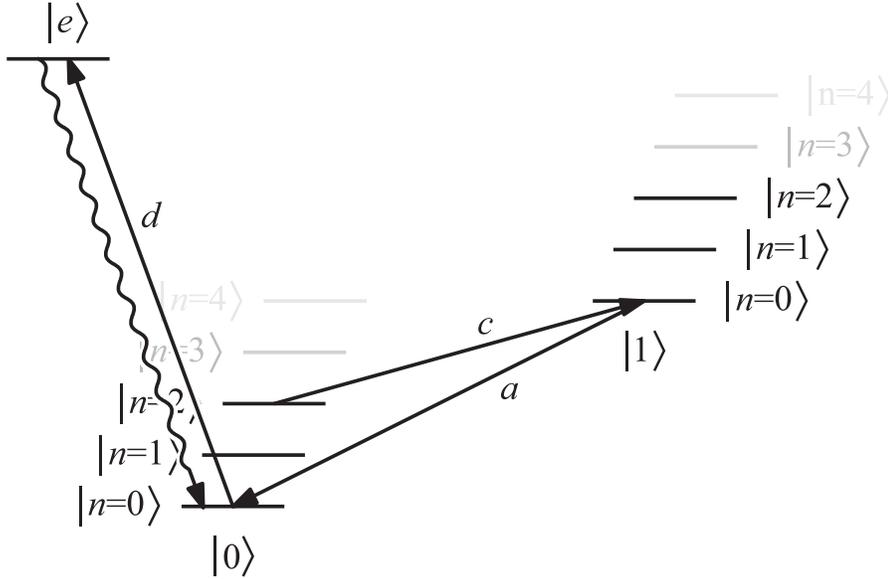}
    \caption{A possible energy level diagram and the transitions used to measure the population
    in the $n=2$ level. If the ion is found in level $\ket{0,n=2}$ after the trap contraction,
    then applying a $\pi$ second red sideband pulse ($c$) followed by a $\pi$ carrier pulse ($a$)
    brings the ion to the state $\ket{0,n=2}$. When the ion is in this latter level, fluorescence can
    be seen by using the detection laser ($d$). On the contrary if
    the ion is initially in $\ket{0,n=0}$ or $\ket{0,n=1}$ the first
    pulse ($c$) would not change the state of the ion since it does
    not have enough energy to be in resonance with the $\ket{1}$
    energy level but the subsequent carrier pulse ($c$) transfer the population to the corresponding
    vibrational levels of $\ket{1}$. Then no fluorescence can be seen when the detection laser ($d$)
    is applied. Note that the transitions $a$ and $c$ should be
    narrow enough not to couple the adjacent vibrational levels and that the detection laser drives a dipole
    transition that couples many of the vibrational levels.}
    \label{fig:particleCreation}
\end{figure}

The simulation can be performed in a system composed of, for
example, two hyperfine states $\ket{0}$ and $\ket{1}$ of an ion. The
states should be coupled by a two-photon Raman transition or by an
RF field in the presence of a magnetic gradient. Notice that the
pairs of Raman beams or the RF field not only couple the two
hyperfine states $\ket{0}$ and $\ket{1}$ but also the vibrational
levels of the ion $\ket{n}=(\ket{n=1}, \ket{n=2}, ...)$. For this
simulation one uses three different coupling fields, one coupled to
the second red sideband (a) $\ket{0, n+2}\rightarrow\ket{1, n}$,
another to the first red sideband (b) $\ket{0,
n+1}\rightarrow\ket{1, n}$ and the last one to the carrier
transition (c) $\ket{0, n}\leftrightarrow\ket{1, n}$. An additional
level $\ket{e}$ is used for fluorescence detection of the population
in level $\ket{0}$. The laser pulse (d) $\ket{0}\rightarrow\ket{e}$
drives a dipole transition that does not resolve the vibrational
levels as the Raman beams or RF field do.

A non-adiabatic increase of the strength of trap potential confining
the ion will simulate the universe expansion. After
non-adiabatically changing the trap potential, the ions motional
state will be a squeezed state where only even (i.e., symmetric with
respect to the trap minimum) vibrational levels contribute. Thus,
the signature of the squeezed state is the population of even
vibrational levels. Measuring the population of the $\ket{n=2}$
vibrational level should be sufficient, since the population in
higher vibrational levels is expected to be negligible. In order to
measure the population of the $n=2$ state one performs a series of
pulses described in figure \ref{fig:particleCreation} that couple
the motional state to the easily readable internal state of the ion.

Since cooling is more efficient for strong confinement, it is
advantageous to begin with a strongly confining potential and cool
to the ground state of motion. Then, after an adiabatic lowering of
the trap potential it is ramped back up non-adiabatically to its
initial strength.

To discriminate against any classical heating, i.e. parametric or
other classical perturbations, the probability of populating the
$n=1$ level can be measured in a similar way as shown in figure
\ref{fig:particleCreation}. The comparison between both
probabilities gives the proof that a squeezed state is generated, if
a higher probability for populating the $n=2$ level than the $n=1$
is obtained.

\subsection{Simulating the Dirac Equation}
\subsubsection{Methods}

Relativistic effects related to the solution of the Dirac equation
can be simulated in trapped ions systems as proposed in
\cite{Bermudez2007} and \cite{Lamata2007}, in particular phenomena
like the {\it Zitterbewegung} (helicoidal motion of the free Dirac
particle as a consequence of the non-commutativity of its velocity
operator components) or  the Klein paradox (transmission of a
relativistic particle of mass $m$ through a potential edge of height
$V>2mc^2$ with nearly no reflection).

In order to simulate the Dirac equation in $3+1$ dimensions for a
spin-1/2 particle in a single trapped ion, L. Lamata {\it et al.}
proposed to design a Hamiltonian that reproduces the same dynamics.
For a free electron in three dimensions the Dirac formalism uses a
bispinor that is a vector valued wave function of four components.
In order to represent all the components one can use a four level
system with states {$\ket{a}$, $\ket{b}$, $\ket{c}$, $\ket{d}$} that
correspond to four internal levels of the ion. The bispinor is
defined as \cite{Lamata2007},
\begin{equation}\label{eq:DiracBispinor}
\ket{\Psi}:=\Psi_a\ket{a}+ \Psi_b\ket{b}+ \Psi_c\ket{c}+\Psi_d
\ket{d}=\left(\begin{array}{c}
\Psi_a\\
\Psi_b\\
\Psi_c\\
\Psi_d
\end{array}
\right)
\end{equation}
%
The dynamic of the system is generated by the pairwise coupling of
the internal levels with the center of mass vibrational levels whose
energies are determined by the trap frequencies $\nu_x$, $\nu_y$,
$\nu_z$. Again, three common types of interaction are used. A
carrier interaction coherently couples two ionic internal levels
without changing the external motion of the ion for example
$\ket{a,n}\leftrightarrow \ket{c,n}$. A Jaynes-Cummings (JC) and an
anti-Jaynes-Cummings (AJC) interactions resonantly couple the ionic
internal with the vibrational levels. The former induce the red
sideband transition and the later the blue one for example
$\ket{a,n+1}\leftrightarrow \ket{c,n}$ and $\ket{a,n}\leftrightarrow
\ket{c,n+1}$ respectively. The corresponding interaction
Hamiltonians under the rotating wave and Lamb-Dicke approximations
are,

\begin{equation}\label{eq:Sidebands-Carrier}
\begin{array}{l}
H_\sigma=\hbar\Omega\left(\sigma^+e^{i\phi}+\sigma^-e^{-i\phi}\right)\\[2 mm]
H_r=\hbar\eta\tilde{\Omega}\left(\sigma^+ae^{i\phi_r}+\sigma^-a^\dag
e^{-i\phi_r}\right)\\[2 mm]
H_b=\hbar\eta\tilde{\Omega}\left(\sigma^+a^\dag
e^{i\phi_b}+\sigma^-a e^{-i\phi_b}\right)\\[2 mm]
\end{array}
\end{equation}

Here again $\sigma^+$ and $\sigma^-$ are the raising and lowering
ionic spin-1/2 operators, $a^\dag$ and $a$ are the creation and
annihilation operators associated with the motional state of the ion
and $\eta=\sqrt{\hbar k^2/2M\nu}$ is the Lamb-Dicke parameter. The
Rabi oscillation frequencies for the blue and red sideband
transitions were set to the same value $\tilde{\Omega}$ and to
$\Omega$ for the carrier transition. \\

When only the carrier coupling is applied, with the appropriate
choice of the phase $\phi$ one obtains,
\begin{equation}\label{eq:Carrier}
H_{\sigma_j}=\hbar\Omega_j \sigma_j
\end{equation}
with $j=\{x, y\}$ and $\sigma_j$ being the Pauli matrices.

When the blue and red sideband couplings are used simultaneously,
the resulting Hamiltonian is the sum of $H_r$ and $H_b$ from
equation \ref{eq:Sidebands-Carrier}. By choosing the phases of the
sideband fields like $\phi_b-\phi_r=\pi$ one obtains,
\begin{equation}\label{eq:RedBlueSid}
\begin{array}{l}
H_{\sigma_j}^p=\pm i\hbar\eta\Delta\tilde{\Omega}\sigma_j
(a-a^\dag).
\end{array}
\end{equation}
Here, $\Delta=\sqrt{\hbar/2M\nu}$ is the spread in position of the
ground wave function. The operator $\sigma_j$ is the Pauli matrix
along the $j=\{x,y\}$ component of the spin. It is not related to
the vibrational directions of motion but to the ionic internal
transitions.

\begin{figure}[tbh]
    \centering
       \includegraphics[width=\columnwidth]{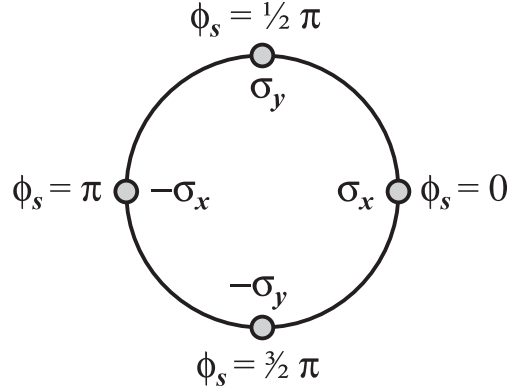}
    \caption{The selection of the phases of the red and blue
    sidebands $\phi_r$ and $\phi_b$ fields are important since they
    shape the form of Hamiltonian. In particular the sum of the
    phases $\phi_s=(\phi_r+\phi_b)/2$ defines the direction of the spin
    operator and also the global sign of the Hamiltonian \ref{eq:RedBlueSid}.
    }
    \label{fig:DiracEquation_phase}
\end{figure}

The selection of the Pauli matrix $\sigma_x=\sigma^++\sigma^-$ or
$\sigma_y=i(\sigma^--\sigma^+)$ as well as the sign of the
Hamiltonian is performed by an additional condition of the applied
red and blue sidebands phases. Even if a condition on their
difference was already used, one needs a supplementary condition to
completely define the phases.  The sum of the phases of the applied
fields $\phi_s=(\phi_b+\phi_r)/2$ determines the sign of the
Hamiltonian, negative for $\phi_s= \{\pi, 3/2\pi\}$ and positive for
$\phi_s= \{0, 1/2\pi\}$ whereas the Pauli matrix $\sigma_j$ is
selected along $j=x$ for $\phi_s= \{0, \pi\}$ and along $j=y$ for
$\phi_s= \{\pi/2, 3/2\pi\}$ (see figure
\ref{fig:DiracEquation_phase}). The phase $\phi_s$ itself does not
modify the dynamics of the system when the previous Hamiltonian is
considered alone, but it is essential to build the Dirac like
Hamiltonian composed of many elements couplings as in equation
\ref{eq:RedBlueSid}. The dynamics of equation \ref{eq:RedBlueSid}
are based on a spin dependent coupling between the internal and
vibrational states that can also be used for the creation of
entangled states of spin and motion. This coupling has already been
applied in previous experiments \cite{Haljan2005,Sackett2000}.

The Hamiltonian of equation \ref{eq:RedBlueSid} constitutes the
basic block to construct the Dirac Hamiltonian in (3+1) dimensions.
Since the spin-1/2 particle to simulate has a momentum with
components in all three spatial directions, one needs to consider
Hamiltonian terms along all of these directions. The operators
$a^\dag$ and $a$ can be associated with the three normal trap
frequencies and therefore with the motion along the three trap axes.
One needs to distinguish $a^\dag_\ell$ and $a_\ell$ for every axis
$\ell={x,y,z}$. The same consideration should be made for the
Lamb-Dicke parameter $\eta_\ell$, the spread in position of the
ground state wave function $\Delta_\ell$ and the Rabi frequency
$\Omega_\ell$ that also need to be defined along the direction of
the applied interaction. The difference between the creation and
annihilation operators can be written in terms of the momentum
operator $p_\ell=i\hbar (a_\ell^\dag-a_\ell)/\Delta_\ell$. Following
the previous considerations, the Hamiltionian of equation
\ref{eq:RedBlueSid} can be rewritten as,
\begin{equation}\label{eq:RedBlueSidFinal}
\begin{array}{l}
H_{\sigma_j}^{p_\ell}=2\eta_\ell\Delta_\ell\tilde{\Omega_\ell}\sigma_j
p_\ell.
\end{array}
\end{equation}

In order to simulate the Dirac dynamics employing four internal
states of an ion, the transitions $\ket{a}\leftrightarrow\ket{c}$,
$\ket{a}\leftrightarrow\ket{d}$, $\ket{b}\leftrightarrow\ket{c}$ and
$\ket{b}\leftrightarrow\ket{d}$ are employed. The corresponding
spin-1/2 operators of equation \ref{eq:RedBlueSidFinal} need to be
defined for the specific transition $\sigma^{ac}_j$,
$\sigma^{ad}_j$, $\sigma^{bc}_j$ and $\sigma^{bd}_j$ and therefore
the corresponding Hamiltonian from equation \ref{eq:RedBlueSidFinal}
can be written as $H_{\sigma_j(ac)}^{p_\ell}$,
$H_{\sigma_j(ad)}^{p_\ell}$, $H_{\sigma_j(bc)}^{p_\ell}$ and
$H_{\sigma_j(bd)}^{p_\ell}$. In the case of the carrier interaction
of equation \ref{eq:Carrier} two internal transitions are used,
giving the following Hamiltonians $H_{\sigma_\ell(ac)}$ and
$H_{\sigma_\ell(bd)}$.

With the correct selection of the direction and phase for the
different fields, one can tailor the desired Hamiltonian as a linear
combination of $H_{\sigma_\ell}$, and $H^{p_\ell}_{\sigma_j}$
yielding, for instance \cite{Lamata2007},
\begin{equation}
\begin{array}{l}
H_D=H_{\sigma_x(ad)}^{p_x}+H_{\sigma_x(bc)}^{p_x}+H_{\sigma_y(ad)}^{p_y}-H_{\sigma_y(bc)}^{p_y}+\\
+H_{\sigma_x(ac)}^{p_z}-H_{\sigma_x(bd)}^{p_z}+H_{\sigma_y(ac)}+H_{\sigma_y(bd)}
\ .
\end{array}
\end{equation}
It should be noticed that all the Rabi oscillation frequencies for
the sideband and carrier transitions $\tilde{\Omega}$ and $\Omega$,
the spread in position $\Delta$ and the Lamb-Dicke parameter $\eta$
are selected equal in all the components of the previous Hamiltonian
for reasons that will be explained later. This means that the fields
employed to drive the various couplings should have the same
relative strength for all the directions of space and the ion trap
should be spherical. Therefore, one may obtain a Hamiltonian $H_D$
that is symmetric with respect to all directions of space. The
Hamiltonian $H_D$ can be written in a $4\times4$ matrix
representation in a base defined by the four internal states of the
ion (see eq. \ref{eq:DiracBispinor}),
\begin{equation}\label{eq:DiracSim}
H_D= \left( \begin{array}{cc} 0 &
2\eta\Delta\tilde{\Omega}(\vec{\sigma}\cdot\vec{p})-i\hbar\Omega \\
2\eta\Delta\tilde{\Omega}(\vec{\sigma}\cdot\vec{p})+i\hbar\Omega & 0
\end{array} \right),
\end{equation}
where each element represents a $2\times2$ matrix. This Hamiltonian
can be compared with the "supersymmetric" representation of the
Dirac Hamiltonian \cite{Thaller1992},
\begin{equation}\label{eq:Dirac3+1}
\textsl{H}_D= \left( \begin{array}{cc} 0 &
c(\vec{\sigma}\cdot\vec{p})-imc^2 \\
c(\vec{\sigma}\cdot\vec{p})+imc^2 & 0
\end{array} \right).
\end{equation}
The speed of light $c$ and the electron rest energy $mc^2$ are
related to the coupling strengths $\Omega$ and $\tilde{\Omega}$
following,
\begin{equation}
c=2\eta\Delta\tilde{\Omega} \quad \mbox{and}\quad mc^2=\hbar\Omega.
\end{equation}

The previous selection of symmetric couplings applied along all the
directions of space represents in a simulation the isotropy of the
velocity of light.

One can recognize the similarities between these two Hamiltonians,
making possible the simulation of a quantum relativistic evolution
of a spin-1/2 particle with an ion-trap experiment.

\begin{figure}[tbh]
    \centering
        \includegraphics[width=\columnwidth]{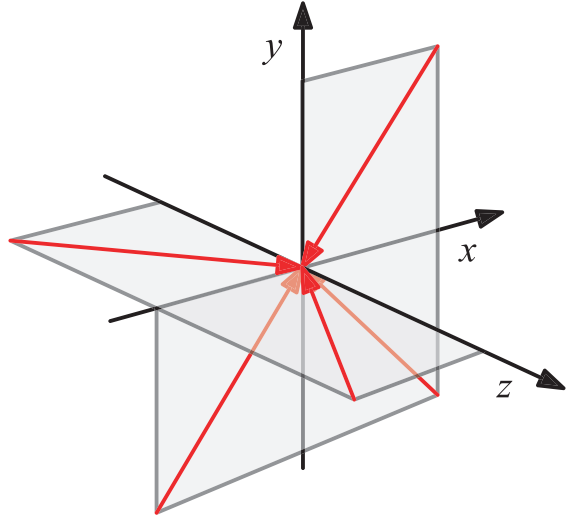}
    \caption{Possible laser configuration to build the Dirac
    Hamiltonian with Raman transitions. In order to build a
    symmetric Hamiltonian as the one of equation
    \ref{eq:DiracSim}, the wave vectors must be the
    same along the three trap axis. For simplification of the lasers requirements
    only one wavelength can be used. Then the four levels
    to perform the simulation need to be close enough in energy i.e. two
    Zeeman split Hyperfine levels as proposed in
    \protect\cite{Lamata2007}.
    Therefore the angles subtended by every pair of Raman beams must be the
    same. Additionally the beams need to be modulated with the
    correct frequency to drive the desired transitions.
    }
    \label{fig:ramanOrientation}
\end{figure}

If the dynamics created by the Hamiltonian of equation
\ref{eq:DiracSim} were reproduced in an experiment with a four level
system using Raman beams, it would require 14 pairs of Raman lasers,
two for each element of the form $H_{\sigma_j}^{p_\ell}$ (see
equation \ref{eq:RedBlueSidFinal})  and one for each carrier element
$H_{\sigma_\ell}$ (see equation \ref{eq:Carrier}). Of course, some
simplifications can be applied, like modulating the frequency of the
beams to generate the blue and the red sideband transitions with a
single pair of beams. But it should be stressed the need of
controlling their phases independently and the minimal requirement
of having at least two pairs of beams (modulated in frequency) with
a resulting k-vector along each one of the trap axes (see figure
\ref{fig:ramanOrientation}).

In order to render the simulation achievable with current
experiments, one can simplify the 3+1 dimensional Hamiltonian of
equation \ref{eq:Dirac3+1} and work with 2+1 dimensions
\cite{Bermudez2007}, or simplify even more to 1+1 dimensions that
still reproduce the most important effects of the Dirac equation.
The resulting 1+1 dimensional Hamiltonian becomes, after a $\pi/2$
rotation around the $x-$axis that changes $\sigma_y$ into $\sigma_z$
\cite{Lamata2007},
\begin{equation}\label{eq:Dirac1+1}
H_D^{(1)}=2\eta\Delta\tilde{\Omega}\sigma_x p_x+\hbar\Omega\sigma_z.
\end{equation}
This Hamiltonian operates on the Dirac "spinor"
$\ket{\Psi^{(1)}}=\Psi^{(1)}_a\ket{a}+ \Psi^{(1)}_b\ket{b}$ build
with a positive ($\Psi^{(1)}_a$) and a negative ($\Psi^{(1)}_b$)
kinetic energy component \cite{Thaller1992}. The term
$2\eta\Delta\tilde{\Omega}\sigma_x p_x$ results from the application
of blue and red sideband fields. It is just the sum of $H_b$ and
$H_r$ with phases $\phi_b=-\pi/2$ and $\phi_r=\pi/2$. The last term
$\hbar\Omega\sigma_z$ can be obtained, for instance, by applying an
off-resonant laser pulse that induces an ac-Stark shift. Because of
the reduction to one spatial dimension, the trapping potential
should also be one-dimensional, that is, a linear trap with strong
radial confinement compared to the axial one would be well suited.

The experimental requirements for simulating the $1+1$ Dirac
Hamiltonian are substantially reduced and achievable in current
experiments.

\subsubsection{Experiment for the simulation of the Zitterbewegung}
By numerically solving the Dirac equation, it is possible to
visualize peculiar behaviour of the probability density function.
One of this peculiarities is the Zitterbewegung where the position
of a particle oscillates even in the absence of external potentials.
Usually this relativistic effect is not seen in experiments since
the frequency is on the order $10^{21}$Hz and the amplitude of order
$10^{-3}${\AA} for a real particle such as an electron.

When simulating the Zitterbewegung with trapped ions the frequency
and amplitude of the oscillatory motion for an ion with average
momentum $p_0$ are given by \cite{Lamata2007} ,
\begin{eqnarray}\label{eq:DiracZitter}
\omega_{\rm ZB}\approx \sqrt{2\eta^2\tilde{\Omega}^2p_0^2/\hbar^2+\Omega^2},\\
R_{\rm
ZB}=\frac{\eta\hbar^2\tilde{\Omega}\Omega\Delta}{4\eta^2\tilde{\Omega}^2\Delta^2p_0^2+\hbar^2\Omega^2}.
\end{eqnarray}
From equation \ref{eq:DiracZitter}, one can obtain a measurable
output with frequencies ranging from zero to some Megahertz and
amplitudes that can go to a thousand Angstroms varying the typical
parameters of Rabi frequencies and trap depths. It can be noticed
that the amplitude of the Zitterwebegung can be made independent of
the Rabi frequencies if $\eta\tilde{\Omega}\approx\Omega$ indicating
that this effect could be seen even with small laser fields.

In order to observe  the Zitterbewegung one needs to determine the
momentum and the position of the ion for different times. In
previous experiments \cite{Meekhof1996A,Meekhof1996B} the state of
motion of the ion was determined by measuring the population
distribution of the Fock states. From the red sideband excitation
probability as a function of time one can fit the different
frequency components associated with the contribution of every
vibrational level. Therefore, the determination of the distribution
over Fock vibrational states requires many data points. Another
possibility could be to measure the Wigner function by instantaneous
measurements, even in the presence of a thermal bath, based on the
proposal in \cite{Santos2007}.

To reduce the number of data points to acquire, one can just measure
the expected value of the generalized quadrature
$Y_\phi=(ae^{-i\phi}-a^\dag e^{i\phi})/2$ as proposed by
\cite{Lougovski2006}. Indeed one does not need to measure the whole
wave function to know the position or momentum of the ion, it is
enough to measure the expected value. The angle $\phi$ is related to
the internal spin state $\ket{+_\phi}=(\ket{a}+e^{i\phi}\ket{b})$
where the ion is initialized. Additionally, a JC interaction needs
to be applied (see $H_r$ in equation \ref{eq:Sidebands-Carrier})
which couples the vibrational state to the internal state where the
measurement is performed.

The probability $P_b$ of finding the ion in the state $\ket{b}$ is
given by (see \cite{Bastin2006}),
\begin{equation}
\begin{array}{lcl}
\displaystyle{\frac{d}{dt}P_b}&=&\displaystyle{\frac{1}{i\hbar}\langle
[\ket{b}\bra{b},H_0+H_r] \rangle + \langle\frac{\partial}{\partial
t}\ket{b}\bra{b}\rangle}\\[2 mm]
               &=&\displaystyle{\frac{1}{i\hbar}\langle [\ket{b},H_r] \rangle}
\end{array}
\end{equation}
The state of the ion at time $t=t_0$ is given by $\rho_{\rm
m}(t_0)\otimes\ket{+_\phi}\bra{+_\phi}$, where $\rho_m$ is density
matrix for the vibrational levels. Then the probability of finding
the ion in state $\ket{b}$ at time $t=t_0$ is given by,
\begin{equation}
\begin{array}{lcl} \displaystyle{\left.
\frac{d}{dt}P_b\right|_{t=t_0}}&=&\displaystyle{\frac{1}{i\hbar}\left.\langle
[\ket{b}\bra{b},H_r] \rangle\right|_{t=t_0}}\\[2 mm]
&=&\displaystyle{{\rm Tr}\left[\rho_{\rm
m}(t_0)\otimes\ket{+_\phi}\bra{+_\phi}[\ket{b}\bra{b},H_r]\right]}\\[2 mm]
&=&\displaystyle{\eta\tilde{\Omega}\langle Y_\phi\rangle.}
\end{array}
\end{equation}

Therefore the generalized quadrature can be written in dimensionless
time  $\tau=\eta {\tilde \Omega}t$ as,
\begin{equation}
\langle Y_\phi\rangle=\left.
\frac{d}{d\tau}P_b^{+_\phi}\right|_{\tau=0}
\end{equation}
When choosing $\phi=-\pi/2$ and $\phi=0$ the position and momentum
operators are measured, respectively. Notice that only two
measurements of the probability of finding the ion in state
$\ket{b}$ are needed to be taken to determine the position or
momentum of the ion.

In order to perform the simulation of the Zitterbewegung the ion can
be cooled to the vibrational ground level, although it is not
required, since it is also possible to measure the Zitterbewegung of
a Doppler cooled ion. Then the internal state of the ion is set into
$\ket{a}$. The next step is to excite the ion to a given motional
state, i.e. a Fock state \cite{Meekhof1996A,Meekhof1996B}, a coherent state
\cite{Carruthers1965}, a thermal state \cite{Stenholm1986}, a
squeezed state \cite{Heinzen1990,Cirac1993}, or an arbitrary state.
The relativistic-like dynamic ruled by the interaction $H_D$ (see
equation \ref{eq:Dirac1+1}) is generated by turning on the fields
resonant with the blue and red sidebands at time $t=0$. The time
evolution of the ion ends at $t=t_0$ when the interaction generated
by the Hamiltonian $H_D$ is turned off. Simultaneously, the internal
state of the ion is driven to $\ket{+_\phi}$ by the application of
an external field, then a JC coupling (see $H_r$ in equation
\ref{eq:Sidebands-Carrier}) is applied to map the ion motional state
into its internal state. At time $t=t_0+\tau$ with
$\tau\ll2\pi/\omega_{ZB}$ the population in the level $\ket{b}$ is
measured by detecting resonance fluorescence. The previous sequence
is repeated with the same parameters for a given number of
repetitions to find the probability $P_b^{+_\phi}(t_0+\tau)$. Then,
the probability $P_b^{+_\phi}$ at $t=t_0$ is also measured (without
the need of applying the JC coupling). From the two last
probabilities the expectation of the generalized quadrature can be
deduced. The whole sequence is then repeated for different $t$ to
extract the trajectory of the ion.

\subsubsection{Experiment of the simulation of the Klein's paradox}
The Klein paradox \cite{Klein1929} is a relativistic effect
described by the Dirac equation. It states that an electron can be
transmitted with almost no reflection through a potential step with
a height $V$ bigger than twice the electron rest energy $mc^2$.

\begin{figure}[tbh]
    \centering
        \includegraphics[width=\columnwidth]{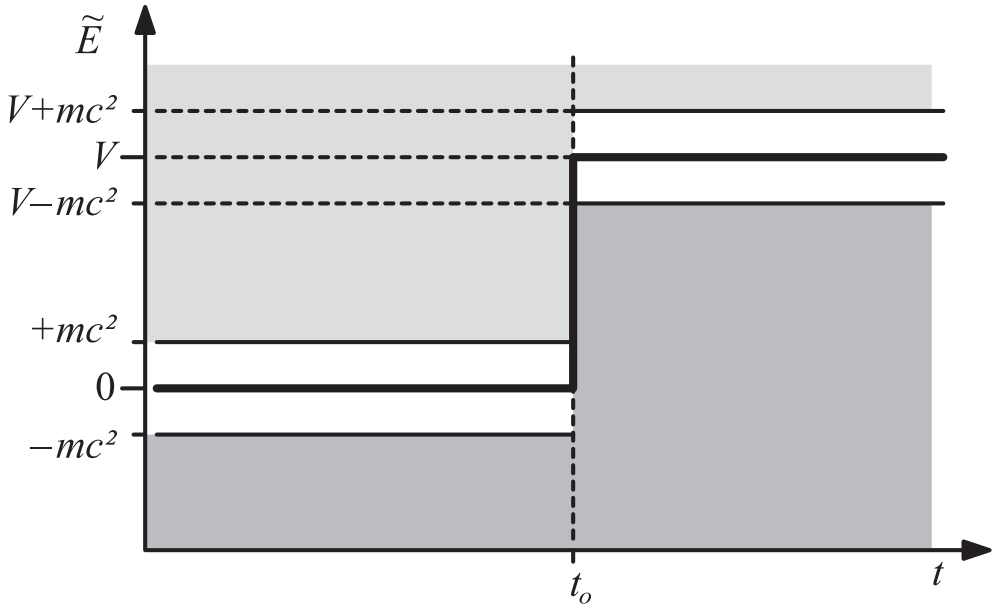}
    \caption{Potential Step (thick line). The white areas represent the forbidden regions
    where the solutions of the Dirac equation are exponentially decaying.
    The light and dark gray areas correspond to the positive and negative energy regions
    respectively. An electron with energy $mc^2<E<V-mc^2$ is found
    initially in the positive energy region. For $t>t_0$, after
    the potential step is risen, the energy lies in the negative
    energy region allowing the creation of a hole in the Dirac sea. Picture adapted from \protect\cite{Schwabl1997}.}
    \label{fig:kleinParadox}
\end{figure}

In the Dirac formalism the particle kinetic energy $E$ takes into
account the rest mass and therefore the electron momentum is given
by
\begin{equation}
p=\frac{1}{c}\sqrt{(E-V)^2-m^2c^4}.
\end{equation}
When $\left|E-V\right|<mc^2$ the momentum is imaginary and the wave
function exponentially decays. However, when $V\geq E+mc^2$, the
momentum is real and one obtains a transmitted wave. This solution
lies in the negative kinetic energy range $E<-mc^2+V$ and therefore
is associated with a electron-positron pair production. The positron
sees an inverted potential step and therefore its motion is not
impeded (see figure \ref{fig:kleinParadox}).

The transmission through a potential step can be reproduced by the
Hamiltonian,
\begin{equation}\label{eq:DiracKlein}
H_V^{(1)}=H_D^{(1)}+V (\ket{a}\bra{a}+\ket{b}\bra{b}).
\end{equation}
The last term is the potential step
$V(\ket{a}\bra{a}+\ket{b}\bra{b})$ that is set at the time $t=t_0$.
It can be implemented for example by the use of a detuned laser that
produces the same stark shift in both levels.

The experimental protocol is summarized in the following steps.
First, the ions are cooled to the ground state $\ket{a}$. Then the
JC and AJC couplings, described in equation \ref{eq:Dirac1+1}, are
applied at $t=0$ by turning on the red and blue sideband excitation
fields. Next, the potential step is  turned rapidly on at time
$t=t_0$. Last, the population in the state $\ket{b}$ is measured by
fluorescence. The probability of finding the ion in the state
$\ket{b}$ is obtained after several repetitions and is equivalent of
the transmission probability through the step potential.

\section*{Acknowledgments}
We thank R. Schmied and E. Solano for stimulating discussions and
comments on the manuscript. We acknowledge financial support by the
European Union (IP QAP and STREP Microtrap), by the Deutsche
Forschungsgemeinschaft and by secunet AG.

\section*{References}
\label{sec:References}

\bibliographystyle{jphysicsB}

\bibliography{qsim}

\end{document}